\documentclass[prd,preprint,superscriptaddress,amsmath,amssymb,nofootinbib]{revtex4}
\usepackage{graphicx}
\usepackage{dcolumn}
\usepackage{bm}
\usepackage{amssymb}
\usepackage{amsmath}
\usepackage{epsfig}    
\usepackage{color}
\usepackage{slashed}
\usepackage{hhline}

\def\be{\begin{equation}}
\def\ee{\end{equation}}
\newcommand{\bea}{\begin{eqnarray}}
\newcommand{\eea}{\end{eqnarray}}

\begin{document}


\title{Scotogenic models with a general lepton flavor dependent $U(1)$ gauge symmetry}

\author{Takaaki Nomura}
\email{nomura@scu.edu.cn}
\affiliation{College of Physics, Sichuan University, Chengdu 610065, China}

\author{Hiroshi Okada}
\email{hiroshi.okada@apctp.org}
\affiliation{Asia Pacific Center for Theoretical Physics (APCTP) - Headquarters San 31, Hyoja-dong,
Nam-gu, Pohang 790-784, Korea}
\affiliation{Department of Physics, Pohang University of Science and Technology, Pohang 37673, Republic of Korea}

\date{\today}

\begin{abstract}
We discuss radiative neutrino mass models with a general lepton flavor dependent $U(1)$ gauge symmetry.
The scotogenic model is adopted for neutrino mass generation in which $Z_2$ odd singlet fermions and an inert scalar doublet are introduced.
A lepton flavor dependent local $U(1)$ symmetry is applied to realize two-zero texture of a Majorana mass matrix of $Z_2$ odd singlet fermions 
where we explore minimal construction choosing $U(1)$ charges of the standard model fermions and singlet fermions. 
Then we investigate neutrino mass matrix and show some predictions for constructed models, and discuss some phenomenological implications.
 \end{abstract}
\maketitle

\section{Introductions}

One of the big mysteries in particle physics is the 
non-zero neutrino masses and their flavor mixings that require physics beyond the standard model (SM). 
Radiative seesaw mechanism is the one of the attractive ways for generating neutrino masses and mixings 
where an active neutrino mass matrix is induced at loop level and the mass value is suppressed by loop factor.
In particular, scotogenic type of radiative neutrino mass models is interesting  since dark matter (DM) candidate often  appears  naturally if we assign
dark $Z_2$ parity to stabilize the DM candidate and forbid tree level neutrino mass; e.g. some earlier works are found in refs.~\cite{Ma:2006km, Kajiyama:2013zla,
Krauss:2002px, Aoki:2008av, Gustafsson:2012vj}).

Besides neutrino mass generation mechanism, introduction of new $U(1)$ gauge (global) symmetry is also an attractive possibility to extend the SM. 
In particular, a lepton flavor dependent $U(1)$ symmetry is interesting since such a symmetry can restrict flavor structure of leptons and would provide interesting phenomenologies.
For example, a predictable neutrino mass model could be realized if we obtain two-zero texture of neutrino mass matrix as a consequence of the symmetry~\cite{Fritzsch:2011qv}.
Thus, we expect such predictive models with a lepton flavor dependent $U(1)$ in which structure of Yukawa couplings associated with neutrino mass generation is restricted by this symmetry; 
some related works can be found in Refs.~\cite{Wang:2019byi,Han:2019diw,Ko:2019tts,Nomura:2018cle,Nomura:2018vfz,Nomura:2017psk,Lee:2017ekw,Ko:2017quv,Kang:2021jmi,Chen:2020jvl,Asai:2019ciz,Araki:2019rmw,Asai:2018ocx,Araki:2012ip,Branco:1988ex,Choubey:2004hn,Baek:2015mna,Crivellin:2015lwa,Plestid:2016esp,Asai:2017ryy,Matsui:2021khj,Nomura:2019dhw,Bhatia:2021eco,Han:2019lux}.
 
In this work we construct scotogenic models with a lepton flavor dependent $U(1)_X$ gauge symmetry that contains $Z_2$ odd SM singlet fermions $N_R$ and an inert doublet $\eta$.
For candidates of $U(1)_X$ we consider general lepton flavor dependent charge as $X = B - x L_e - y L_\mu - z L_\tau$, with $B$ and $L_i$ being Baryon and lepton number ($i=e,\mu,\tau$), 
where $x+y+z =3$ is required to cancel anomalies~\footnote{It can be obtained by linear combinations of anomaly free $U(1)_{B-L}$ and $U(1)_{L_i - L_j}$ symmetries.}; a few studies of type-I seesaw models with this kind of gauge symmetry are found in ref.~\cite{Araki:2012ip, Asai:2019ciz}.
In constructing models, we adopt the following criteria for choosing the $U(1)_X$ charges and field contents; (1) Yukawa couplings of the terms $\overline{L_L} \eta N_R$ and $\overline{L_L} H e_R$ are diagonal, 
(2) the Majorana mass matrix of $N_R$ has two-zero structure, (3) only one singlet scalar $\varphi$ has to break $U(1)_X$ for minimality, (4) all three generations of leptons have non-zero $U(1)_X$ charge.
We then formulate active neutrino mass matrix induced at one loop level and make numerical analysis to search for allowed parameter sets.

This paper is organized as follows.
In Sec. II, we introduce our models with lepton flavor dependent $U(1)$ gauge symmetry, choosing charge assignment for the SM fermions, and provide a formula of neutrino mass matrix. 
 In Sec.III we carry out numerical analysis for neutrino mass matrix and search for allowed parameter sets that accommodate neutrino oscillation data, and discuss phenomenological implications of the models. 
 Finally, we conclude several results in Sec. IV.

\section{Model setup}
\begin{table}[t!]
\begin{tabular}{|c||c|c|c|c|c|c|c|c|c|}\hline\hline  
& ~$Q_{L_i}$~&~$u_{R_i}$~&~$d_{R_i}$~&~$L_{L_i}$~& ~$e_{R_i}$~& ~$N_{R_i}$~ & ~$H$~&~$\eta$~& ~$\varphi$~ \\\hline\hline 
$SU(3)_c$   & $\bm{3}$  & $\bm{3}$  & $\bm{3}$ & $\bm{1}$ & $\bm{1}$  & $\bm{1}$ & $\bm{1}$ & $\bm{1}$  & $\bm{1}$     \\\hline 
$SU(2)_L$   & $\bm{2}$  & $\bm{1}$  & $\bm{1}$ & $\bm{2}$ & $\bm{1}$  & $\bm{1}$ & $\bm{2}$ & $\bm{2}$  & $\bm{1}$     \\\hline 
$U(1)_Y$    & $\frac16$  & $\frac23$ & $-\frac13$ & $-\frac12$ & $-1$  & $0$ & $1/2$ & $-1/2$ & $0$ \\\hline
$U(1)_X$    & $\frac13$  & $\frac13$ & $\frac13$ & $\{-x, -y, -z \}$ & $\{-x, -y, -z \}$  & $\{-x, -y, -z \}$ & $0$ & $0$ & $2$ \\\hline
\end{tabular}
\caption{Charge assignments of the quarks, leptons and scalar fields
under $SU(3)_c \times SU(2)_L\times U(1)_Y \times U(1)_X$ ($X= B- xL_e - y L_\mu - z L_\tau$), where the lower index $i$ is the number of family that runs over $1-3$ and $Z_2$ odd is imposed for $N_R$ and $\eta$. 
To cancel anomaly $\{x, y, z \}$ should satisfy $x+y+z =3$. }\label{tab:1}
\end{table}

In this section, we introduce models that are based on a general lepton flavor dependent $U(1)_X \ (X= B- xL_e - y L_\mu - z L_\tau)$ gauge symmetry.
To cancel anomalies $\{x,y,z\}$ should satisfy $x+y+z = 3$ where $x=y=z=1$ corresponds to well known $U(1)_{B-L}$ case.
We choose $x$ and $y$ as free parameter and $z = 3 -x -y$.
A singlet scalar field with $U(1)_X$ charge 2 is introduced to break the $U(1)_X$ spontaneously by its vacuum expectation value (VEV).
We also introduce a $SU(2)$ doublet scalar $\eta$ with hypercharge $Y=-1/2$ and singlet fermions $N_R$ that are odd under $Z_2$ symmetry; the other fields are all $Z_2$ even.
All the fields in the model are summarized in Tab.~\ref{tab:1} showing their charge assignments.
The scalar potential is written by
\begin{align}
V = & \mu_H^2 H^\dagger H + \mu_\eta^2 \eta^\dagger \eta + \mu_\varphi^2 \varphi^* \varphi + \lambda_H (H^\dagger H)^2 + \lambda_\eta (\eta^\dagger \eta)^2 + \lambda_\varphi (\varphi^* \varphi)^2 \nonumber \\
& + \lambda_{H \eta} (H^\dagger H)(\eta^\dagger \eta) 
+ {\lambda'}_{H \eta} (H \eta)(\eta H)  
+ \tilde{\lambda}_{H \eta} (H \eta)^2  
+ \lambda_{H \varphi} (H^\dagger H)(\varphi^* \varphi)
+ \lambda_{\varphi \eta} (\varphi^* \varphi)(\eta^\dagger \eta).
\end{align}
Also the Lagrangian for relevant Yukawa interactions of the model is
\begin{align}
\mathcal{L}_Y = y_\ell \overline{L_L} e_R H + y_\eta \overline{L_L} N_R \eta + y_N \overline{N_R^c} N_R \varphi + \tilde{y}_N \overline{N_R^c} N_R \varphi^* + h.c. ,
\end{align}
where $c$ indicates charge conjugation.
Note that non-zero components of these Yukawa couplings are determined by our choice of $U(1)_X$ charges for fermions.

\subsection{Scalar and gauge boson masses}
We require the SM Higgs field and singlet $\varphi$ develop their VEVs to break electroweak and $U(1)_X$ symmetry spontaneously.
These fields are written by
\begin{equation}
H =\left[\begin{array}{c}
G^+\\
\frac{v + \tilde{h} +i G_Z}{\sqrt2}
\end{array}\right],\quad 
\varphi_{1,2}=
\frac{v_{\varphi}+ \phi + i G_{Z'} }{\sqrt2},
\end{equation}
where $G^+$ and $G_Z$ are massless Nambu-Goldstone(NG) bosons which are absorbed by the SM gauge bosons $W^+$ and $Z$ respectively, and
 $G_{Z'}$ corresponds to NG boson absorbed by extra neutral gauge boson $Z'$ from $U(1)_{X}$.  
The VEVs are obtained by solving stationary conditions $\frac{\partial {\cal V}}{\partial v} = \frac{\partial {\cal V}}{\partial v_\varphi} =0$ which are explicitly written by
\begin{align}
& v \left( \mu_H^2 + \frac{\lambda_{H \varphi}}{2} v_\varphi^2 \right) + \lambda_H v^3 =0, \\
& v_\varphi \left( \mu_\varphi^2 + \frac{\lambda_{H \varphi}}{2} v^2  \right) + \lambda_\varphi v_\varphi^3 =0.
\end{align}
After the scalar fields developing VEVs, neutral scalars $\tilde h$ and $\phi$ mix each other.
In the analysis of this work, we assume the mixing between $\tilde h$ and $\phi$ is small choosing $\lambda_{H \varphi}$ to be small, for simplicity, and $\tilde h$ is identified as the SM Higgs boson.

We require the $Z_2$ odd scalar $\eta$ not to develop VEV so that $Z_2$ symmetry is not broken.
The real and imaginary part of neutral component in the inert scalar $\eta$ obtain different masses after symmetry breaking.
From the scalar potential we find the masses as 
\begin{equation}
m_{\eta_R} = \sqrt{\mu_\eta^2 + \frac{\lambda_{H\eta}+{\lambda'}_{H \eta}}{2} v^2 + \frac{\tilde \lambda_{H\eta} }{2} v^2 }, \quad m_{\eta_I} = \sqrt{\mu_\eta^2 + \frac{\lambda_{H\eta}+{\lambda'}_{H \eta}}{2} v^2 - \frac{\tilde \lambda_{H\eta} }{2} v^2 }.
\end{equation}
Thus the mass difference is given by the parameter $\tilde \lambda_{H \eta}$ as
\begin{equation}
\delta m^2_\eta \equiv m^2_{\eta_R} - m^2_{\eta_I} = \frac{\tilde \lambda_{H \eta}}{2} v^2.
\end{equation}
 The charged component in $\eta$ has mass $m_{\eta^\pm} = \sqrt{\mu_\eta^2 + (\lambda_{H \eta}+{\lambda'}_{H \eta}) v^2/2}$.

The $U(1)_X$ gauge symmetry is spontaneously broken by the VEV of $\varphi$.
Then $Z'$ boson from $U(1)_X$ obtains its mass that is given by
\begin{equation}
m_{Z'} = 2 g_X v_\varphi,
\end{equation}
where $g_X$ is gauge coupling constant associated with $U(1)_X$.

\subsection{Majorana mass matrix of $N_R$}

The relevant terms associated with Majorana mass of $N_R$ are given by
\begin{equation}
\mathcal{L} \supset \frac12 \tilde M_N \overline{N_R^c} N_R +  y_N \overline{N_R^c} N_R \varphi +  \tilde{y}_N \overline{N_R^c} N_R \varphi^{*} + h.c. \, ,
\end{equation}
where the first term is bare Majorana mass term of $N_R$.
Non-zero components of these terms are determined by the choice of $N_R$ charges, i.e. values of $\{x, y\}$.
After $\varphi$ developing its VEV, the Majorana mass term for $N_R$ is given by
\begin{equation}
\frac12 \left( \tilde{M}_N + \sqrt{2} (y_N + \tilde{y}_N) v_\varphi \right) \overline{N_R^c} N_R = \frac12 M_N \overline{N_R^c} N_R.
\label{eq:MR}
\end{equation}
The mass matrix is diagonalized as $V^T M_N V$ where $V$ is a unitary matrix. 
The mass eigenstate is given by $\psi_R = V N_R$.

We consider several models choosing the values of $x$ and $y$ that determine the structure of Yukawa couplings and Majorana mass matrix of $N_R$.
The $U(1)_X$ charge structures of relevant Yukawa interactions and bare Majorana mass term are given by
\begin{align}
& \left( \overline{L_L} e_R H \right)_{\rm U(1)_X \, charge} = \left( \overline{L_L} N_R \eta \right)_{\rm U(1)_X \, charge} = \left( \begin{array}{ccc} 0 & x-y & x-z \\ -x+y & 0 & y-z \\ -x+z & -y+z & 0 \end{array} \right), \\
& \left( \overline{N_R^c} N_R \varphi^{(*)} \right)_{\rm U(1)_X \, charge} = \left( \begin{array}{ccc} -2x \pm 2 & -x-y \pm 2 & -x-z \pm 2 \\ -x-y \pm 2 & -2y \pm 2 & -y-z \pm 2 \\ -x-z \pm 2 & -y-z \pm 2 & -2z \pm 2 \end{array}  \right), \\
& \left( \overline{N_R^c} N_R \right)_{\rm U(1)_X \, charge} =  \left( \begin{array}{ccc} -2x & -x-y  & -x-z  \\ -x-y  & -2y  & -y-z  \\ -x-z  & -y-z  & -2z  \end{array}  \right), 
\end{align}
where $+2$ is for $\varphi$ and $-2$ is for $\varphi^*$ at the RHS of the second equation.
We choose $x$ and $y$ to obtain simple structure for Yukawa coupling and Majorana mass for realizing high predictability in neutrino mass generation.
The adapted criteria are as follows:
\begin{itemize}
\item $x \neq 0$ and $y \neq 0$,
\item Yukawa coupling matrices $y_\ell$ and $y_\eta$ are diagonal,
\item The Majorana mass matrix $M_N$ in Eq.~\eqref{eq:MR} has two-zero texture.
\end{itemize}
For the second criterion, we require $x \neq y \neq z$.
We then find six possible scenarios which realize two-zero texture of $M_R$, which are summarized as follows.
\begin{align}
&(1) \ \{x,y\} = \{-1,1 \} \to M_N : \left( \begin{array}{ccc} X & X & X \\ X & X & 0 \\ X & 0 & 0 \end{array} \right), \quad
(2) \ \{x,y\} = \{1,-1 \} \to M_N : \left( \begin{array}{ccc} X & X & 0 \\ X & X & X \\ 0 & X & 0 \end{array} \right), \nonumber \\
&(3) \ \{x,y\} = \{3, 1 \} \to M_N : \left( \begin{array}{ccc} 0 & 0 & X \\ 0 & X & X \\ X & X & X \end{array} \right), \quad
(4) \ \{x,y\} = \{3,-1 \} \to M_N : \left( \begin{array}{ccc} 0 & X & 0 \\ X & X & X \\ 0 & X & X \end{array} \right), \nonumber \\
&(5) \ \{x,y\} = \{1, 3 \} \to M_N : \left( \begin{array}{ccc} X & 0 & X \\ 0 & 0 & X \\ X & X & X \end{array} \right), \quad
(6) \ \{x,y\} = \{-1,3 \} \to M_N : \left( \begin{array}{ccc} X & X & X \\ X & 0 & 0 \\ X & 0 & X \end{array} \right),
\end{align}
where $X$ indicates a non-zero component of $M_N$ and $z$ is given by $3 -x -y$~\footnote{We can identify these cases are linear combinations of $U(1)_{B - 3 L_i}$ and $U(1)_{L_j - L_k}$.}.

\subsection{Neutrino masses}

\begin{figure}[tb]
\begin{center}
\includegraphics[width=100.0mm]{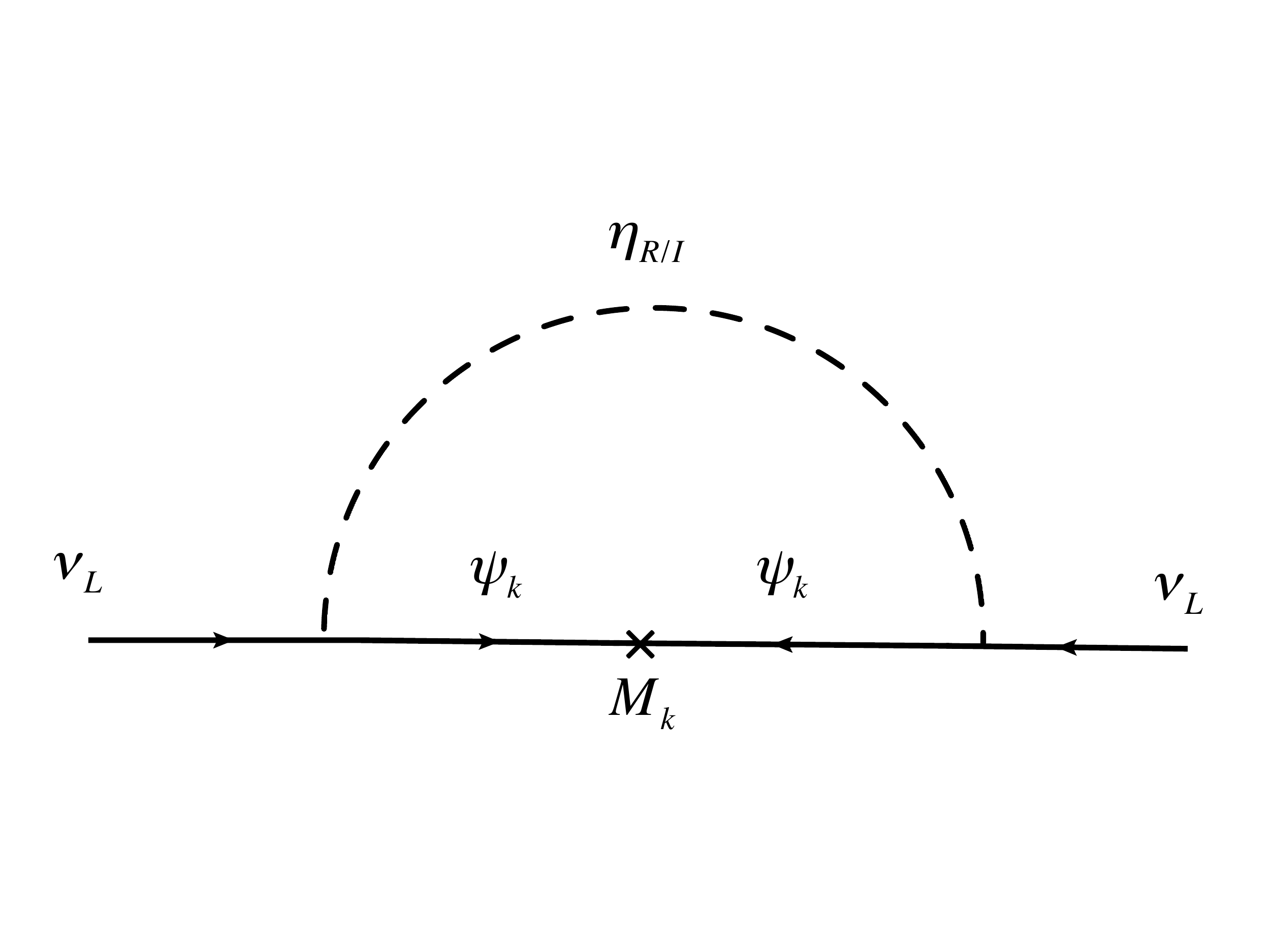} 
\caption{One loop diagram generating active neutrino mass matrix. }
  \label{fig:diagram}
\end{center}\end{figure}

The relevant interaction term for neutrino mass generation is 
\begin{equation}
\mathcal{L}_Y \supset  \frac{1}{\sqrt{2}} \tilde y_\eta \overline{L_L} \psi_R (\eta_R + i \eta_I) + h.c., 
\end{equation}
where $\tilde y_\eta = y_\eta V$.
We obtain the neutrino mass matrix at one-loop level by diagram in Fig.~\ref{fig:diagram} such that 
\begin{equation}
(m_\nu)_{ij} = \frac{ (\tilde y_\eta^*)_{ki} (\tilde y_\eta^*)_{kj}}{32 \pi^2} m_{\psi_k} \left[ \frac{m_{\eta_R}^2 \ln \left( \frac{m_{\eta_R}^2}{m_{\psi_k}^2} \right)}{m_{\eta_R}^2 - m_{\psi_k}^2}
- \frac{m_{\eta_I}^2 \ln \left( \frac{m_{\eta_I}^2}{m_{\psi_k}^2} \right)}{m_{\eta_I}^2 - m_{\psi_k}^2} \right]. \label{eq:neutrino-mass}
\end{equation}
We diagonalize the matrix $m_\nu$ by a unitary matrix $U_{\rm PMNS}$; $U_{\rm PMNS}^T m_\nu U_{\rm PMNS}\equiv {\rm diag}(m_1,m_2,m_3)$.  

%
Several observables regarding neutrino physics are as follows. 
Firstly, we write the mass square differences	
\begin{align}
(\mathrm{NH}):\  \Delta m_{\rm atm}^2 =m_3^2 - m_1^2,
\quad
(\mathrm{IH}):\    \Delta m_{\rm atm}^2 =m_2^2 - m_3^2,
 \end{align}
where $\Delta m_{\rm atm}^2$ is atmospheric neutrino mass square difference, and NH and IH indicate the normal hierarchy and the inverted hierarchy, respectively. 
Solar mass square difference is also represented by
\begin{align}
\Delta m_{\rm sol}^2=m_2^2 - m_1^2.
 \end{align}
We estimate the value and compare it with experimental data.
 %
The PMNS matrix $U_{\rm PMNS}$ is parametrized by three mixing angles $\theta_{ij} (i,j=1,2,3; i < j)$, one CP violating Dirac phase $\delta_{CP}$,
and two Majorana phases $\{\alpha_{21}, \alpha_{32}\}$ such that
\begin{equation}
U_{\rm PMNS}= 
\begin{pmatrix} c_{12} c_{13} & s_{12} c_{13} & s_{13} e^{-i \delta_{CP}} \\ 
-s_{12} c_{23} - c_{12} s_{23} s_{13} e^{i \delta_{CP}} & c_{12} c_{23} - s_{12} s_{23} s_{13} e^{i \delta_{CP}} & s_{23} c_{13} \\
s_{12} s_{23} - c_{12} c_{23} s_{13} e^{i \delta_{CP}} & -c_{12} s_{23} - s_{12} c_{23} s_{13} e^{i \delta_{CP}} & c_{23} c_{13} 
\end{pmatrix}
\begin{pmatrix} 1 & 0 & 0 \\ 0 & e^{i \frac{\alpha_{21}}{2}} & 0 \\ 0 & 0 & e^{i \frac{\alpha_{31}}{2}} \end{pmatrix},
\end{equation}
where $c_{ij}$ and $s_{ij}$ stands for $\cos \theta_{ij}$ and $\sin \theta_{ij}$ respectively. 
The mixing angles can be derived in terms of the components of $U_{\mathrm{PMNS}}$ as follows:
\begin{align}
\sin^2\theta_{13}=|(U_{\mathrm{PMNS}})_{13}|^2,\quad 
\sin^2\theta_{23}=\frac{|(U_{\mathrm{PMNS}})_{23}|^2}{1-|(U_{\mathrm{PMNS}})_{13}|^2},\quad 
\sin^2\theta_{12}=\frac{|(U_{\mathrm{PMNS}})_{12}|^2}{1-|(U_{\mathrm{PMNS}})_{13}|^2}.
\end{align}
In addition, we can compute the Jarlskog invariant and $\delta_{CP}$ from PMNS matrix elements $U_{\alpha i}$:
\begin{equation}
J_{CP} = \text{Im} [U_{e1} U_{\mu 2} U_{e 2}^* U_{\mu 1}^*] = s_{23} c_{23} s_{12} c_{12} s_{13} c^2_{13} \sin \delta_{CP}.
\end{equation}
The Majorana phases can be also estimated in terms of other invariants $I_1$ and $I_2$:
\begin{equation}
I_1 = \text{Im}[U^*_{e1} U_{e2}] = c_{12} s_{12} c_{13}^2 \sin \left( \frac{\alpha_{21}}{2} \right), \
I_2 = \text{Im}[U^*_{e1} U_{e3}] = c_{12} s_{13} c_{13} \sin \left( \frac{\alpha_{31}}{2} - \delta_{CP} \right).
\end{equation}
Moreover, the effective mass for the neutrinoless double beta decay is given by
\begin{align}
\langle m_{ee}\rangle=  |m_1 \cos^2\theta_{12} \cos^2\theta_{13}+m_2 \sin^2\theta_{12} \cos^2\theta_{13}e^{i\alpha_{21}}+m_3 \sin^2\theta_{13}e^{i(\alpha_{31}-2\delta_{CP})}|,
\end{align}
where the predicted value can be compared with the current constraints; for example the strongest constraint is given by  KamLAND-Zen~\cite{KamLAND-Zen:2022tow}. 

\section{Numerical analysis and phenomenology}

In this section, we carry out numerical analysis to fit the neutrino data and find some predictions of the models.
Then some implications for phenomenology are discussed.

\subsection{Numerical analysis}

\begin{figure}[tb]
\begin{center}
\includegraphics[width=70.0mm]{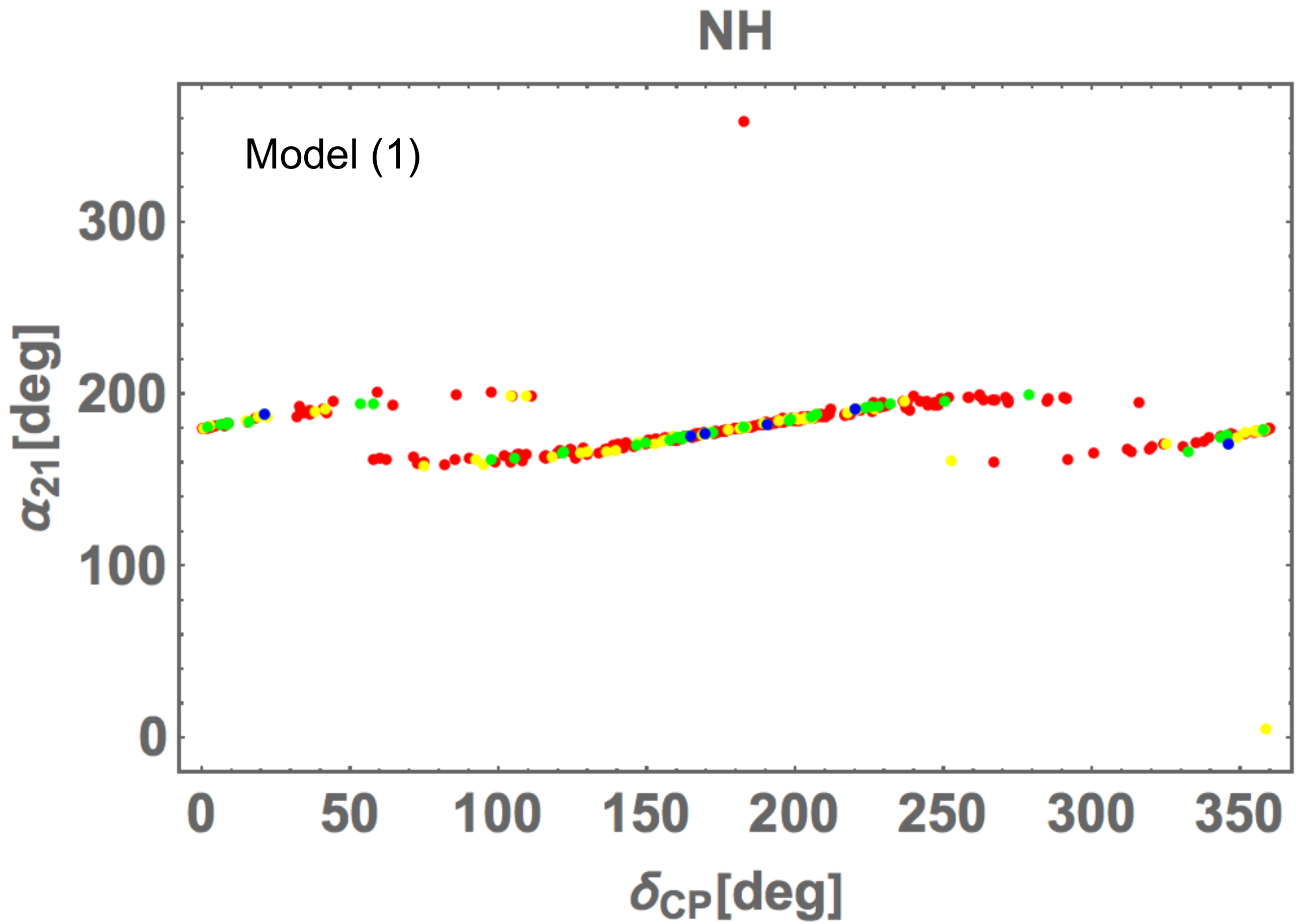} \quad
\includegraphics[width=70.0mm]{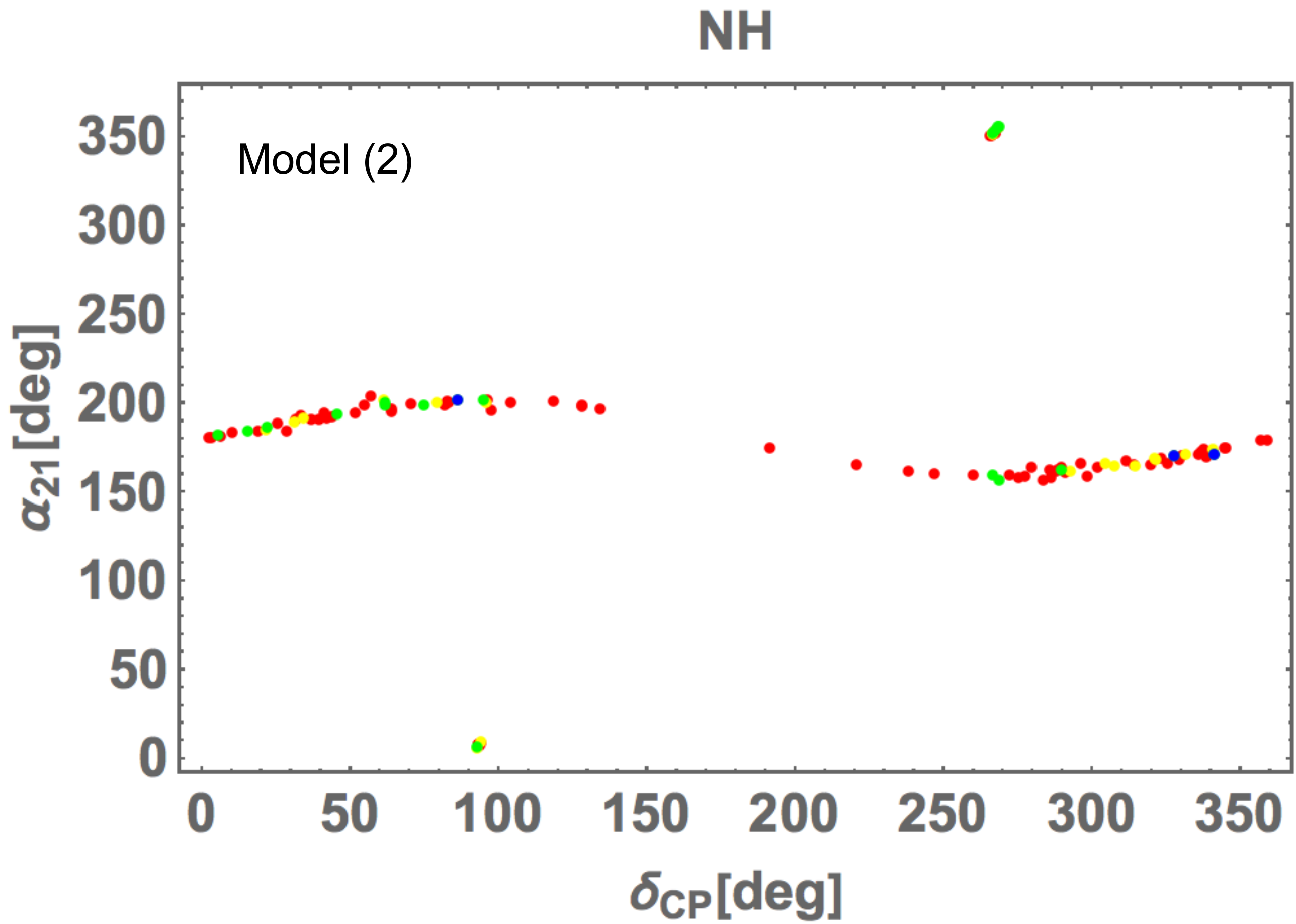} 
\includegraphics[width=70.0mm]{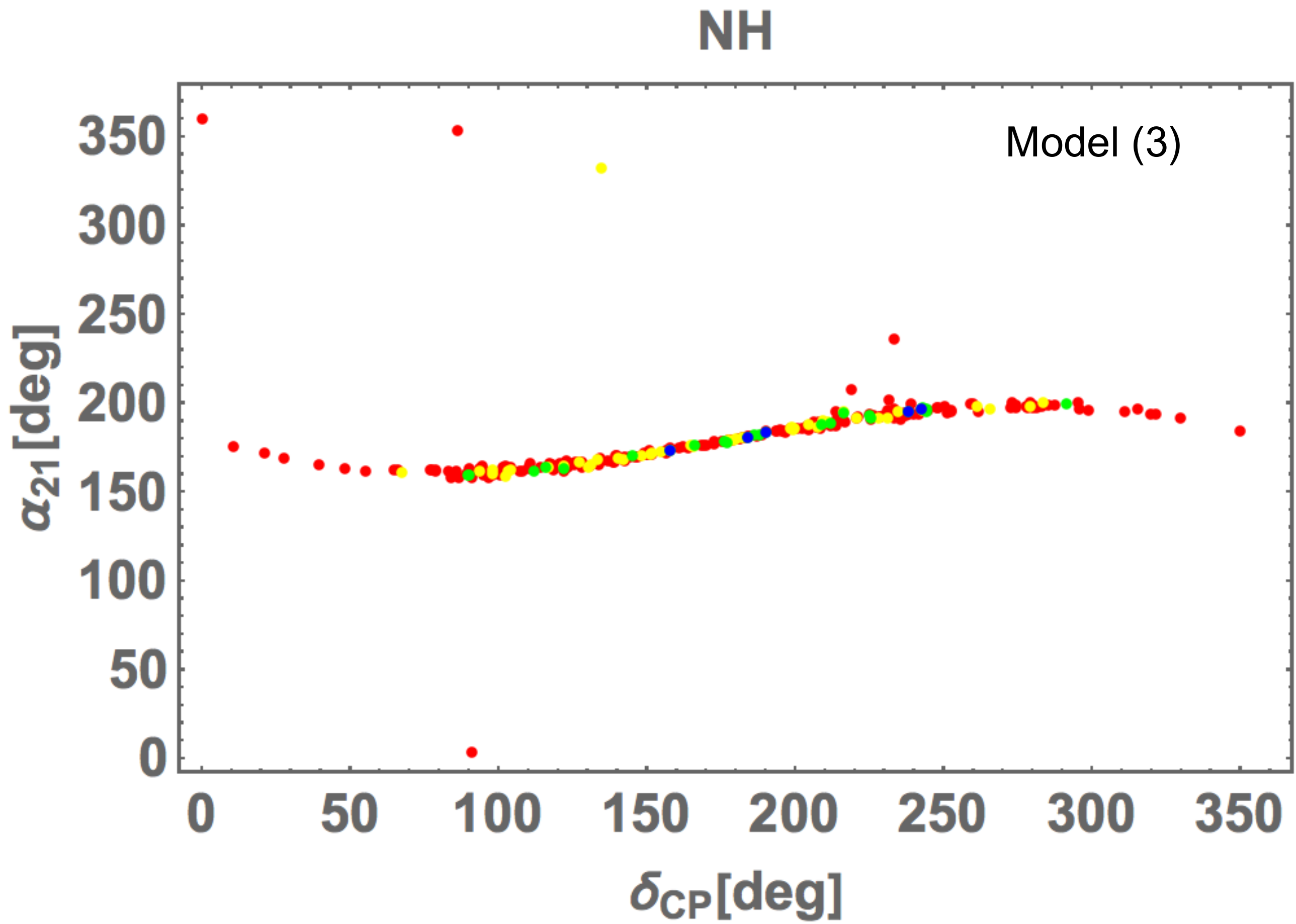} \quad
\includegraphics[width=70.0mm]{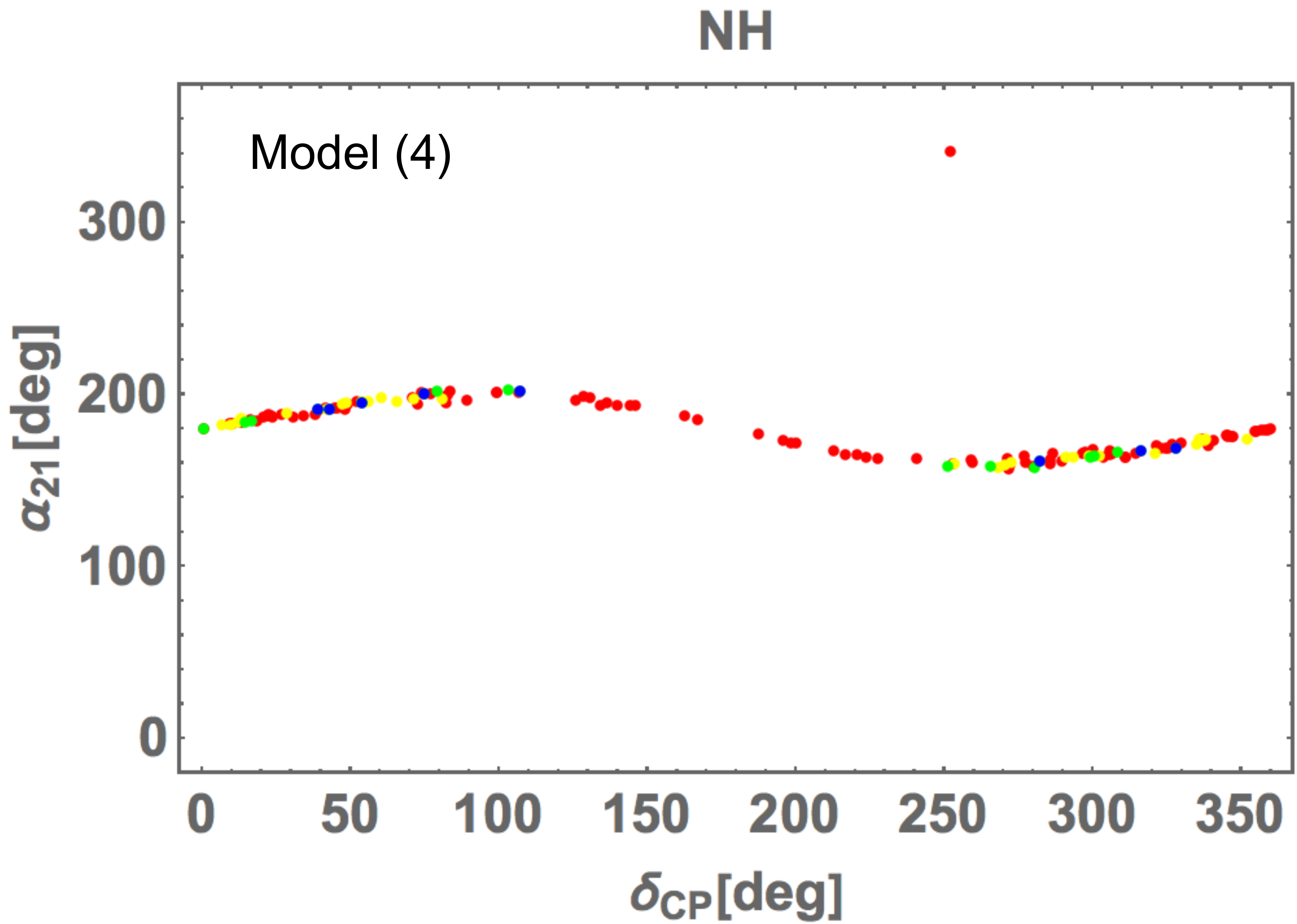} 
\includegraphics[width=70.0mm]{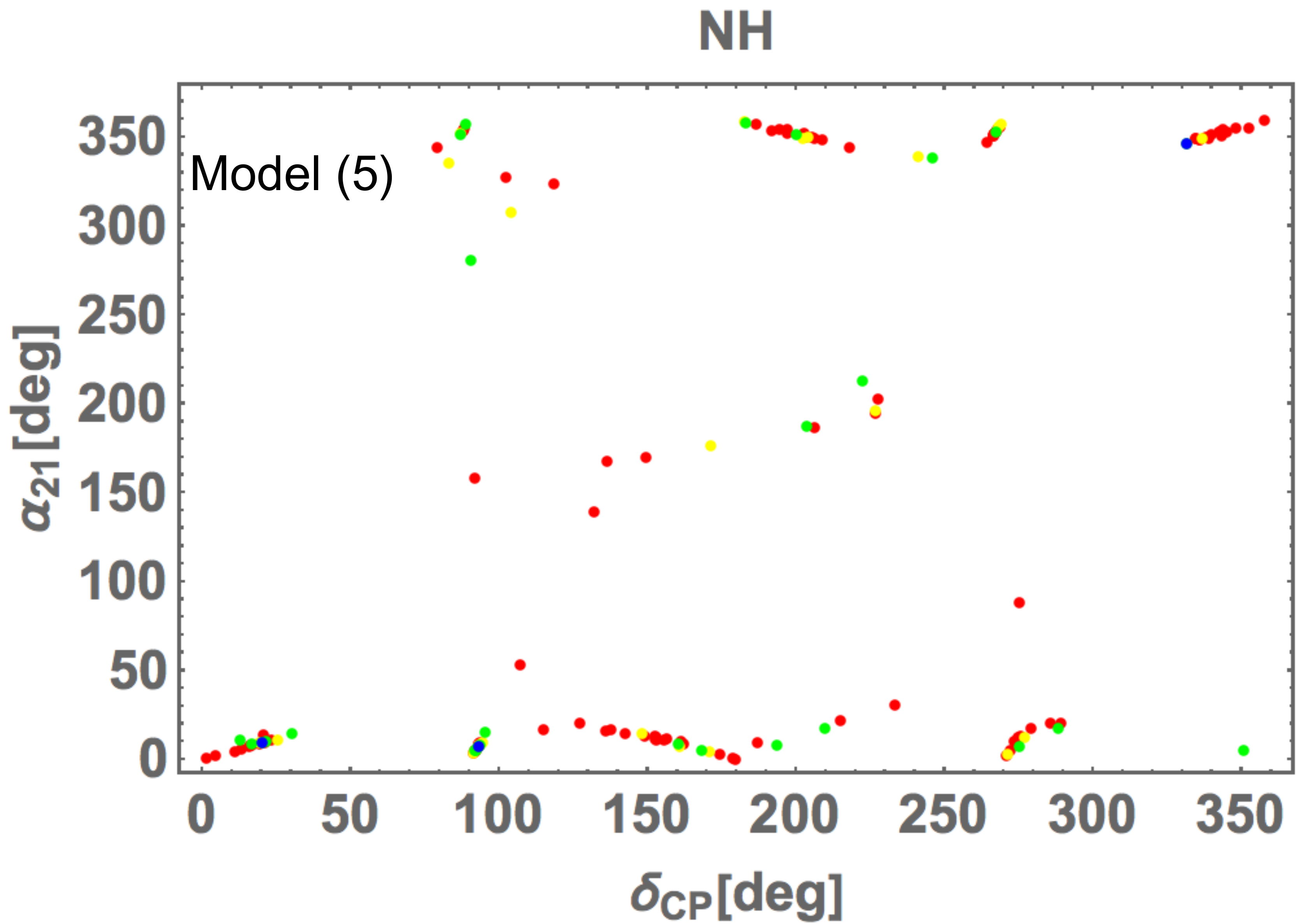} \quad
\includegraphics[width=70.0mm]{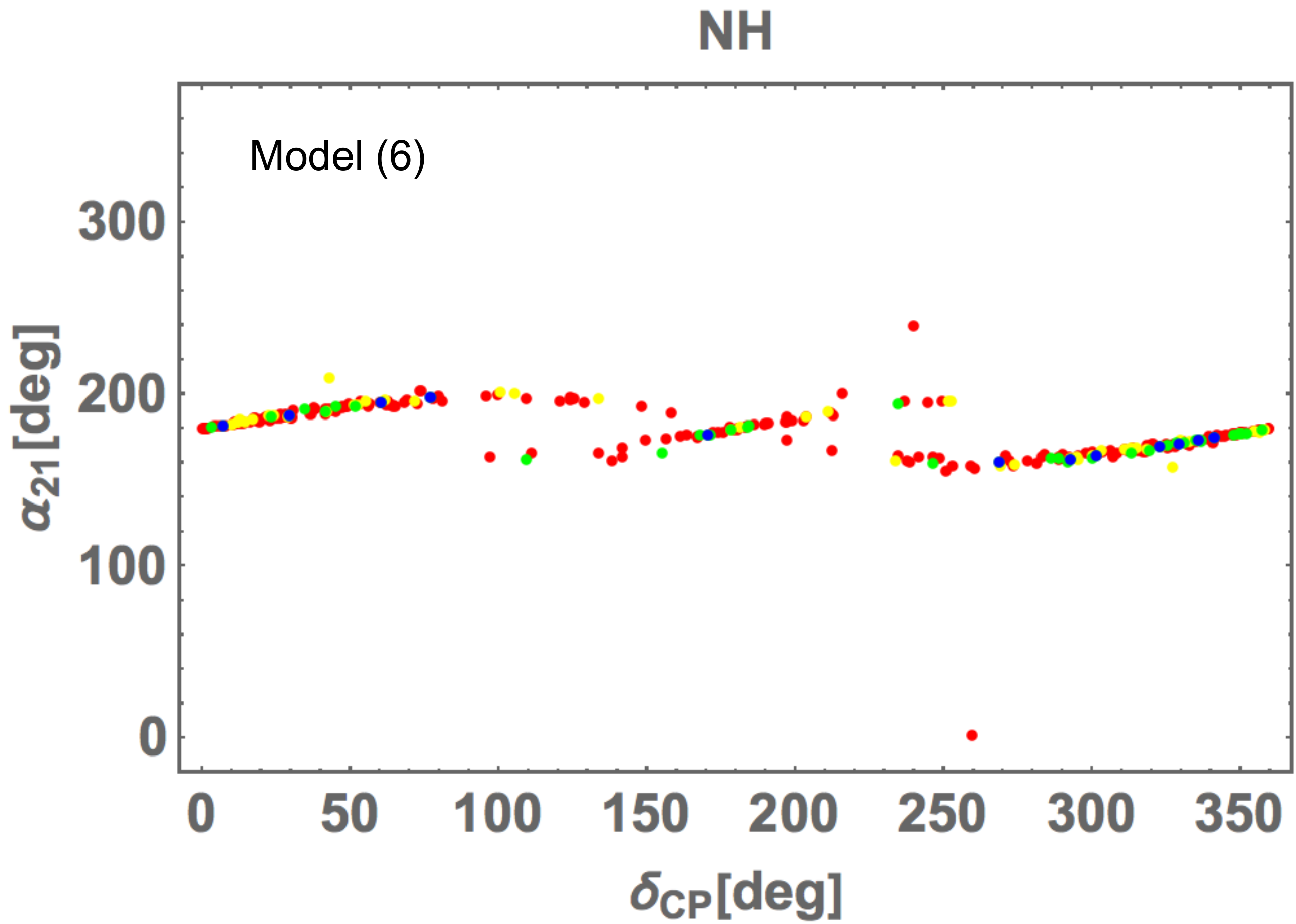} 
\caption{The correlations between $\delta_{CP}$ and $\alpha_{21}$ in NH case for each model. The blue, green, yellow, and red color points correspond to $\sigma\le1$, $1<\sigma\le2$, $2<\sigma\le3$, and $3<\sigma\le5$ interval, respectively in $\chi^2$ analysis. }
  \label{fig:phaseNH}
\end{center}\end{figure}

\begin{figure}[tb]
\begin{center}
\includegraphics[width=70.0mm]{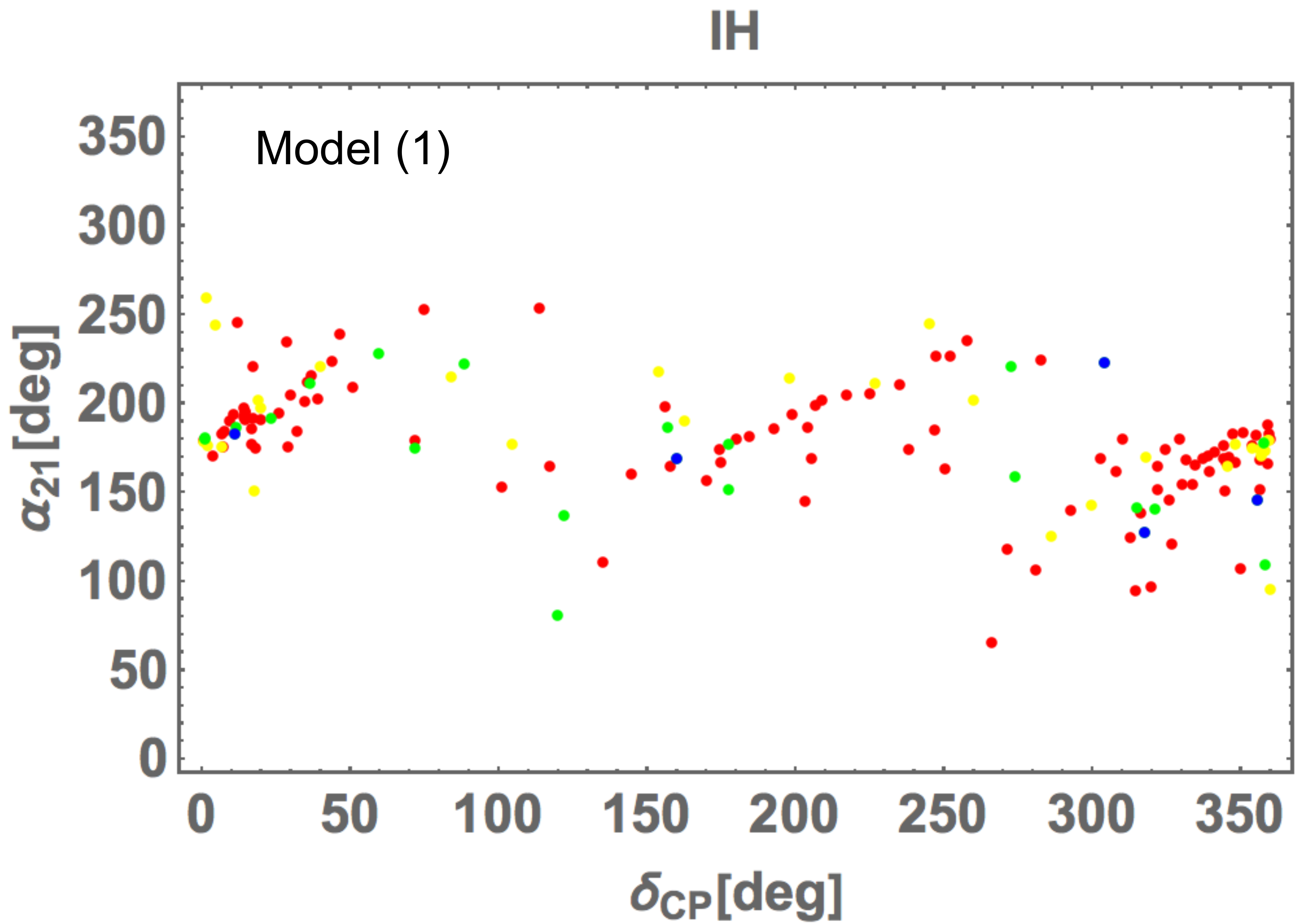} \quad
\includegraphics[width=70.0mm]{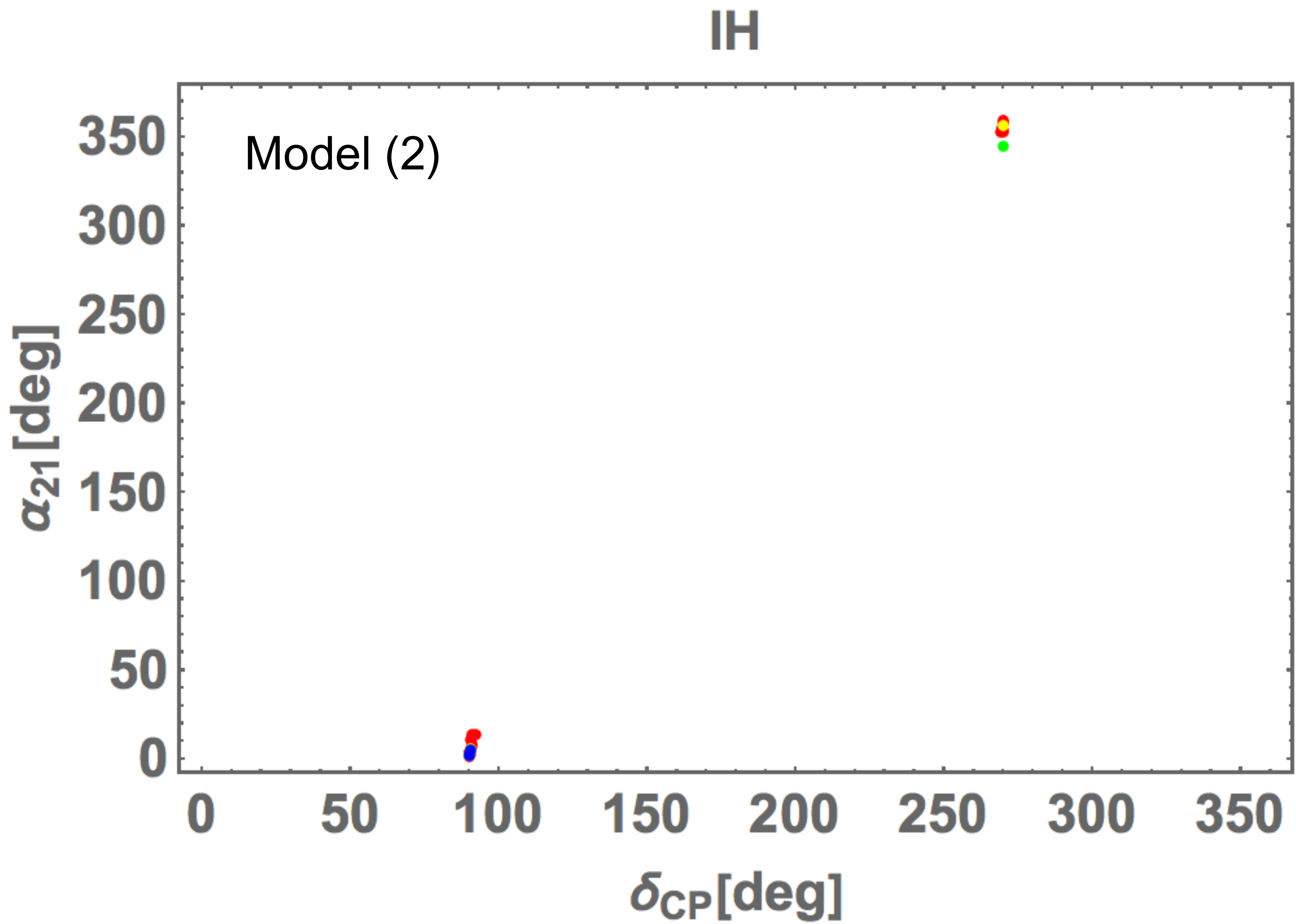} 
\includegraphics[width=70.0mm]{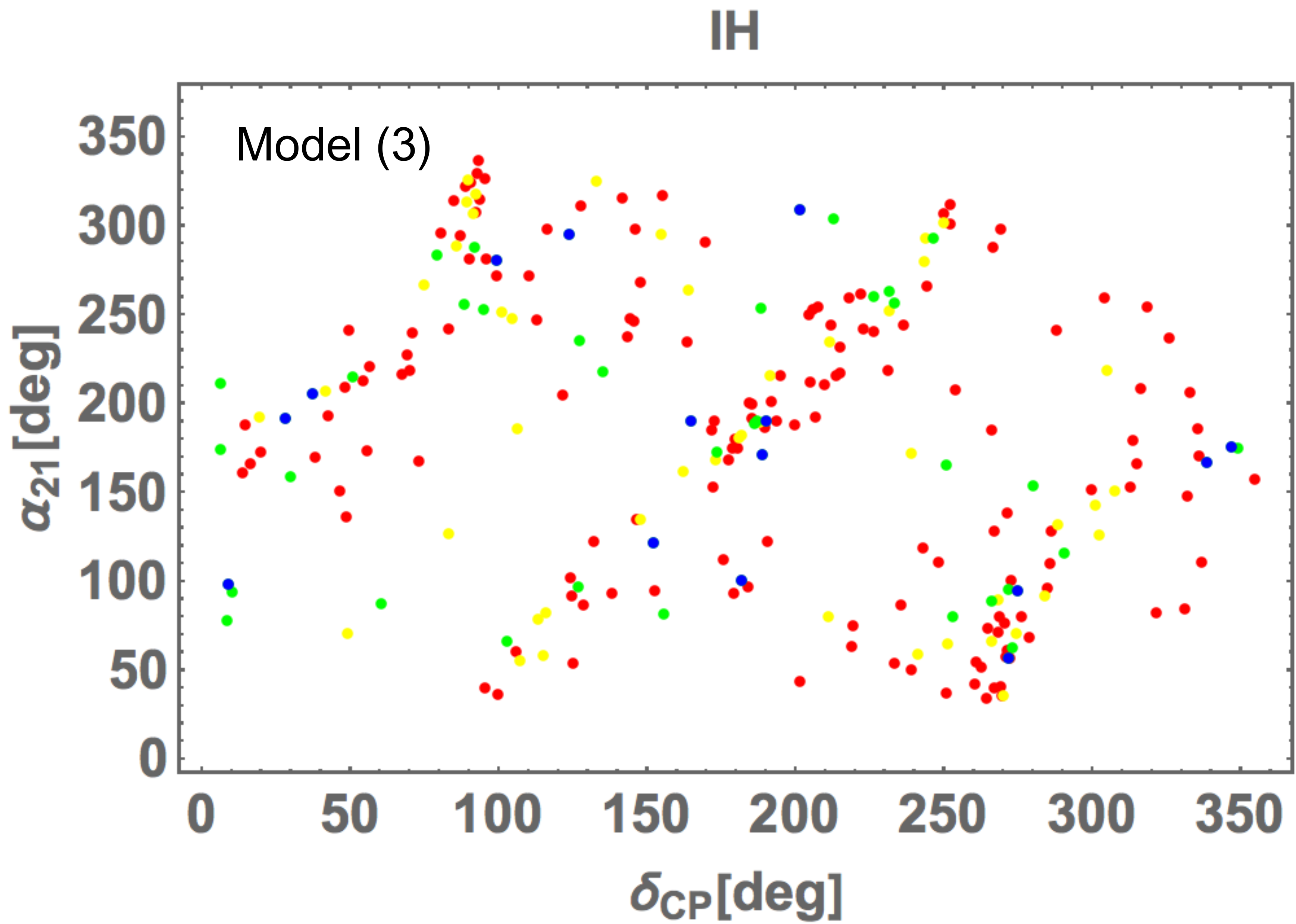} \quad
\includegraphics[width=70.0mm]{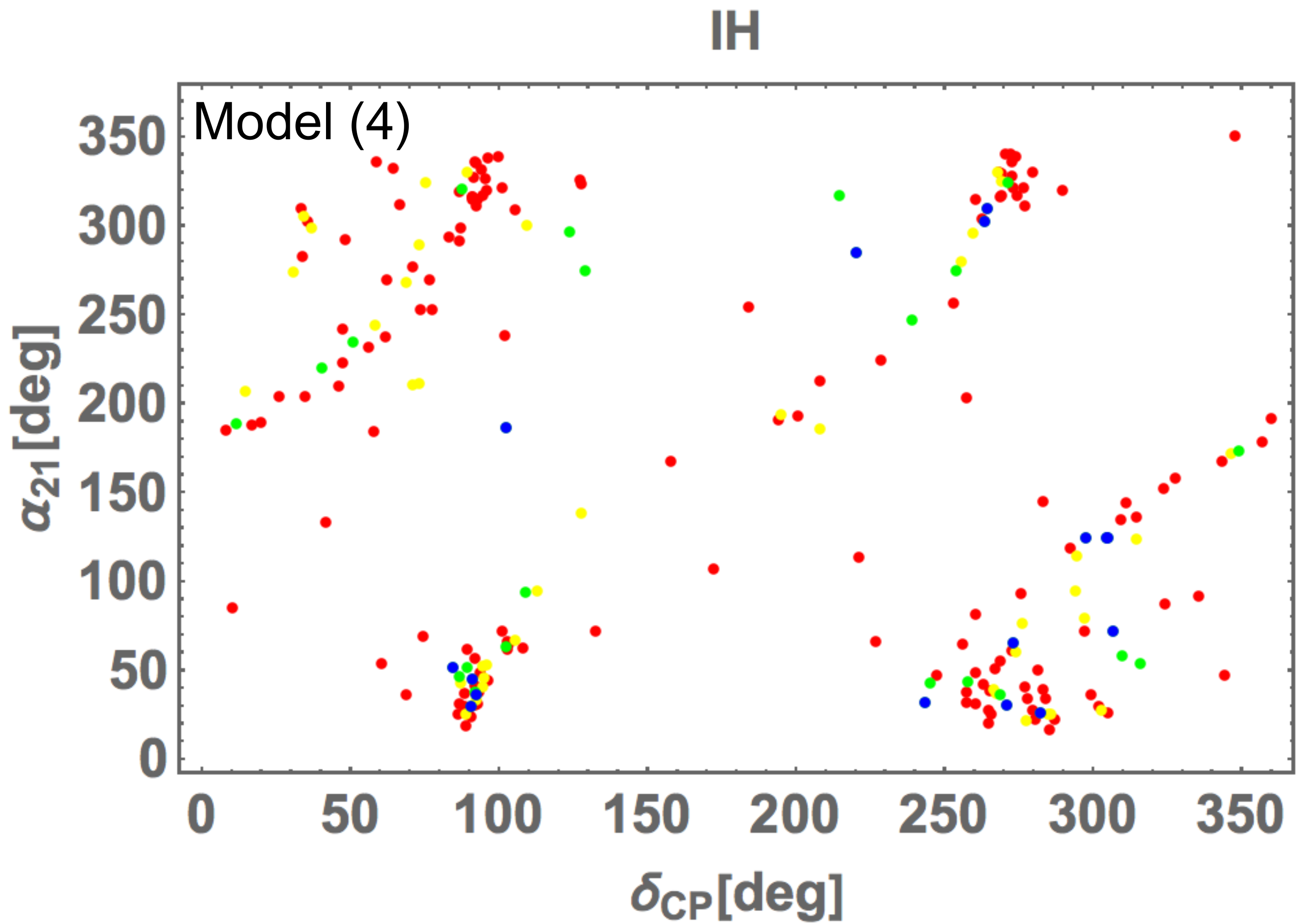} 
\includegraphics[width=70.0mm]{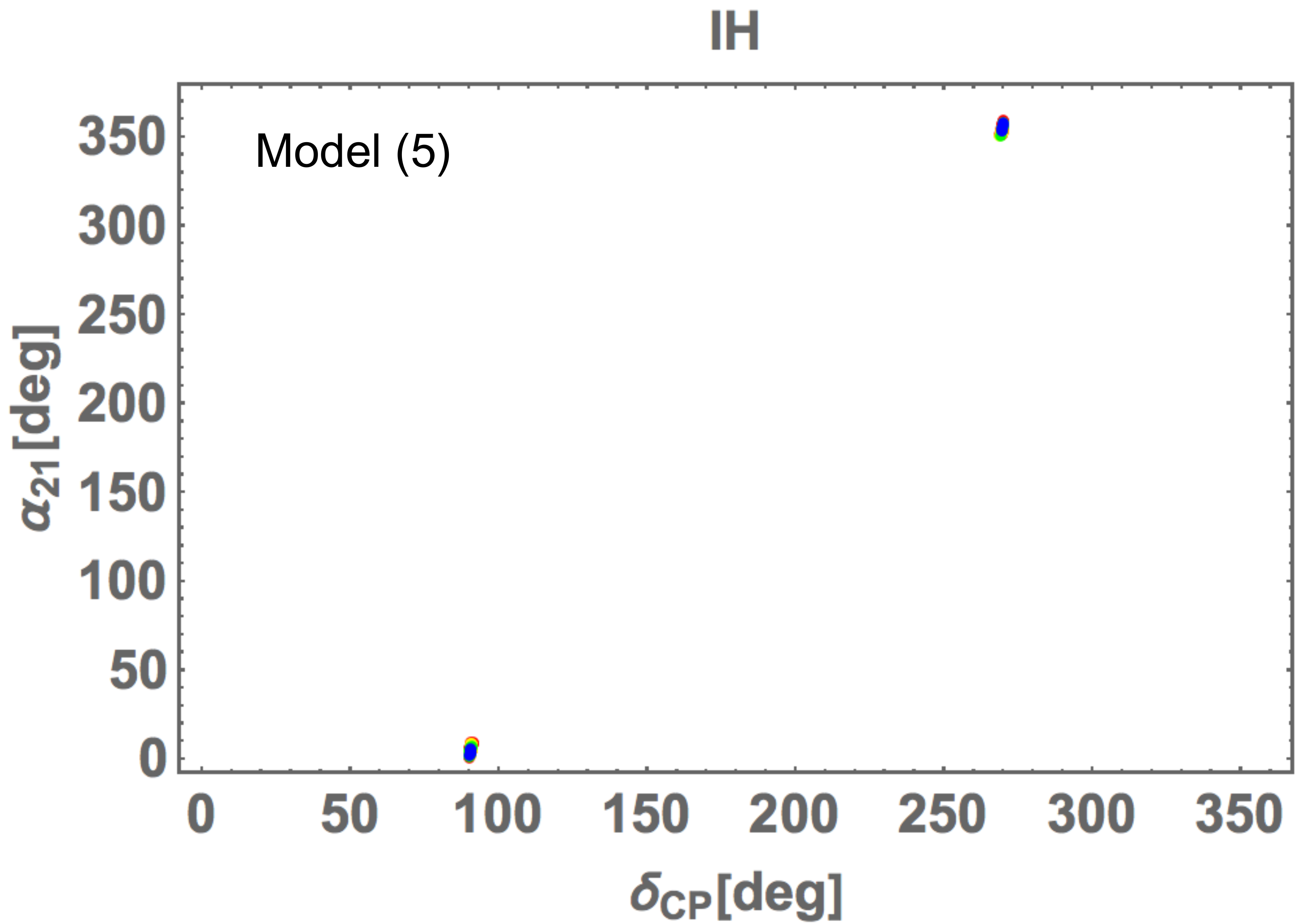} \quad
\caption{The correlations between $\delta_{CP}$ and $\alpha_{21}$ in IH case for each model. The colors of points are the same as Fig.~\ref{fig:phaseNH}; it is the same in following plots.  }
  \label{fig:phaseIH}
\end{center}\end{figure}

The relevant free parameters for neutrino mass are Yukawa coupling $y_{\eta_i}$, the elements of Majorana mass matrix $m_{ij} \equiv (M_N)_{ij}$, inert scalar boson mass $m_{\eta_R}$ and mass difference $\delta m_\eta^2 = m_{\eta_R}^2 - m_{\eta_I}^2$.
{Note that we choose $y_{\eta_i}$ and one of the non-zero element of $m_{ij}$ to be real that can be achieved by phase redefinition of fermions without loss of generality.}
For Yukawa coupling we factor out $y_1$ as $(y_1, y_2, y_3) = y_1 (1, \hat{y}_2, \hat{y}_3)$, and $y_1$ is chosen to fit $\Delta m^2_{atm}$.
Then we scan relevant free parameters in the range of  
\begin{equation}
\{ |m_{ij}| , m_{\eta_R} \} \in [10^2, 10^5] \ {\rm GeV}, \quad \{ \hat{y}_2,  \hat{y}_3 \} \in [10^{-3}, 10^2],  \quad \delta m_\eta^2 \in [10^{-2}, 1] v^2,
\end{equation}
where perturbativity condition of Yukawa coupling, $y_i \lesssim \sqrt{4 \pi}$, is also imposed.
For each parameter point, neutrino masses and mixings are estimated using Eq.~\eqref{eq:neutrino-mass},
and $\chi^2$ value is calculated using neutrino data from NuFit 5.2~\cite{Esteban:2020cvm} for $\{\sin \theta_{12}, \sin \theta_{23}. \sin \theta_{13}, \Delta m^2_{\rm atm}, \Delta m_{\rm sol}\}$.
We then search for the parameter sets which realize $\chi^2$ value giving $\sigma \lesssim 5$ confidence level and explore some predictions regarding neutrino measurements such as CP phases, $\langle m_{ee} \rangle$ and $\sum m_i$.
As a result we find possible parameter sets except for IH case in model (6).

In Fig.~\ref{fig:phaseNH}, we show the correlation between $\delta_{CP}$ and $\alpha_{21}$ in NH case for each model.
The blue, green, yellow, and red color points correspond to $\sigma\le1$, $1<\sigma\le2$, $2<\sigma\le3$, and $3<\sigma\le5$ interval, respectively in $\chi^2$ analysis.
For models except for (5), the Majorana phase $\alpha_{21}$ tends to be around [150, 200] [deg] with a few points outside the region where we find some specific patterns for the correlation between $\alpha_{21}$ and $\delta_{CP}$.
In model (5), we have more points giving $\alpha_{21}$ around [0, 30] [deg] and [320, 360] [deg], and few points  around [150, 200] [deg] where correlation between the phases is less clear.
In Fig.~\ref{fig:phaseIH}, we also show the correlation between $\delta_{CP}$ and $\alpha_{21}$ in IH case for each model.
For model (1), we find $\alpha_{21}$ tends to be around [100, 250] [deg] and the correlation between the phases is not clear.
For model (2) and (5), we find specific prediction regarding $\delta_{CP}$ and $\alpha_{21}$ where these values are localized at around $\{\alpha_{21}, \delta_{CP} \} = \{5, 90\} $[deg] and $\{270, 355 \}$[deg].
On the other hand, for model (3) and (4), we find the possible range of the phases are wide and their correlations are not strong.

\begin{figure}[tb]
\begin{center}
\includegraphics[width=70.0mm]{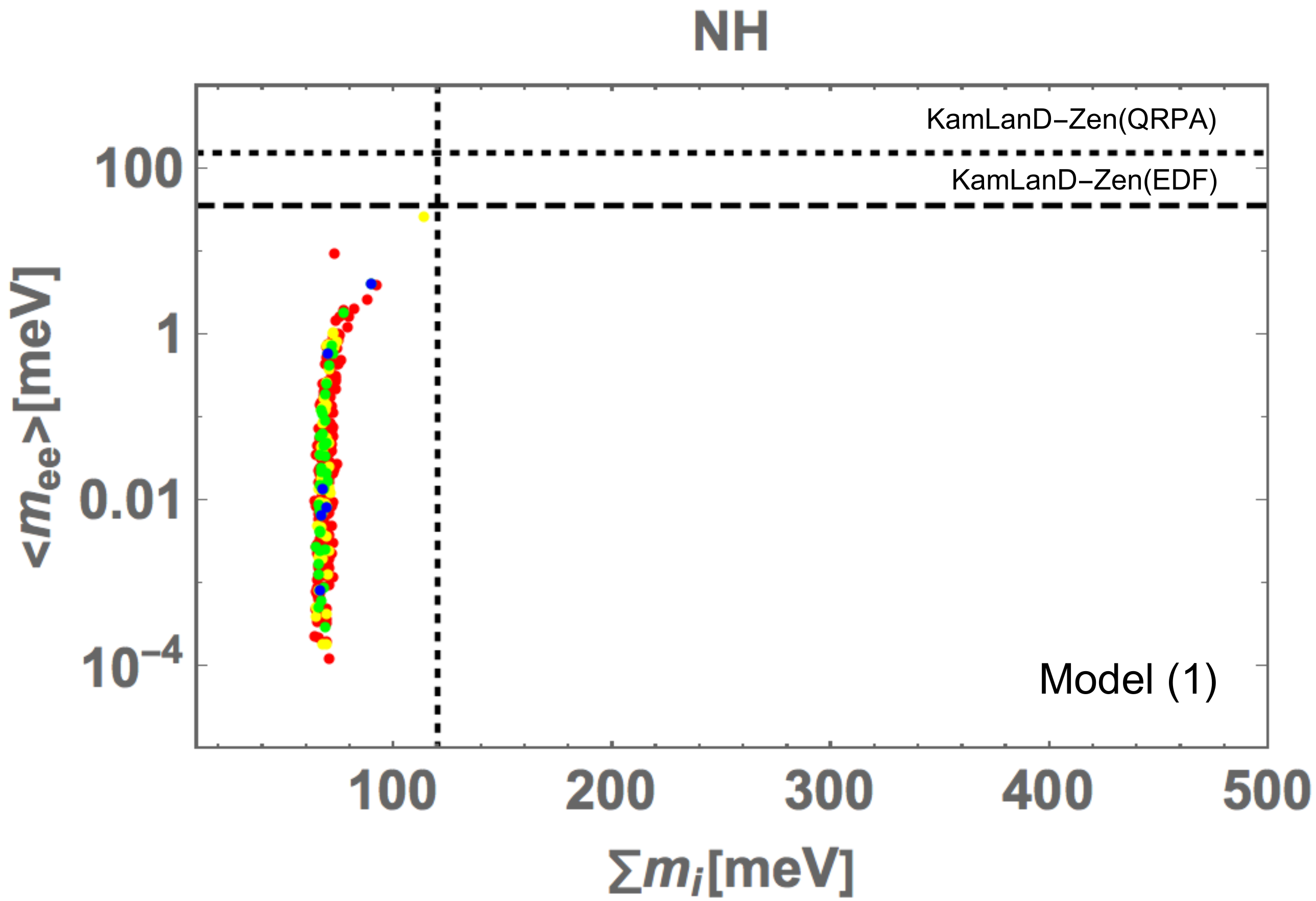} \quad
\includegraphics[width=70.0mm]{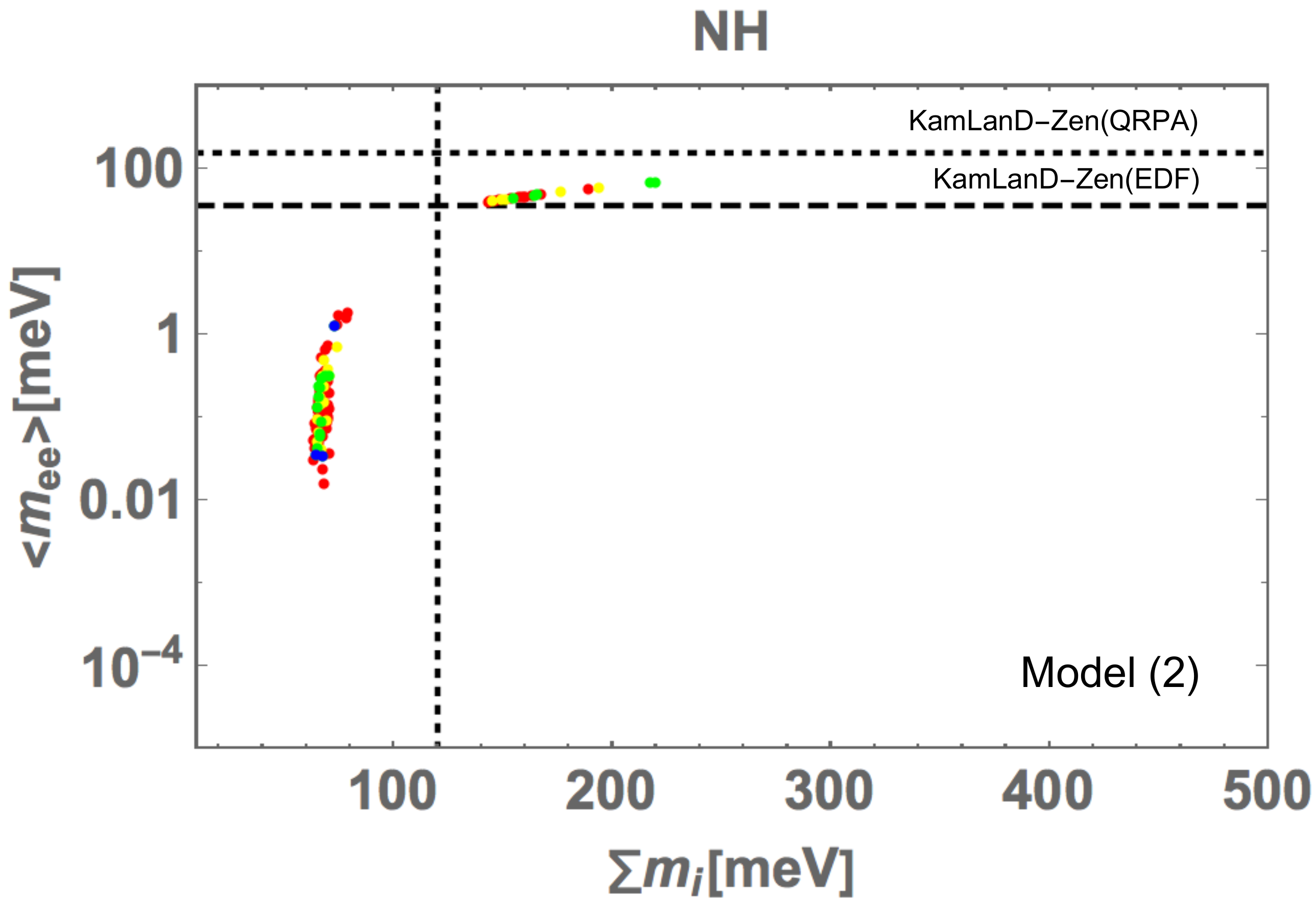} 
\includegraphics[width=70.0mm]{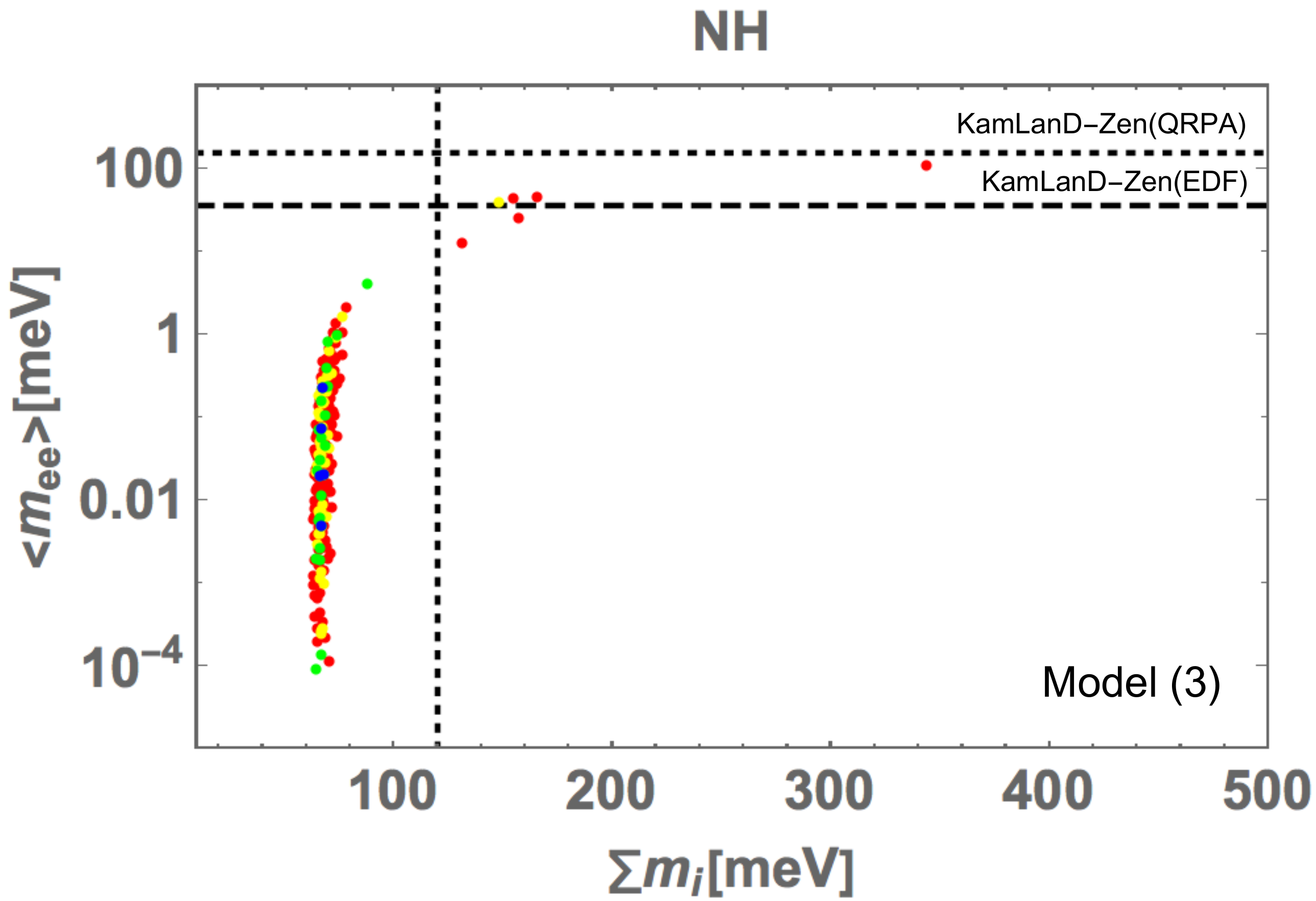} \quad
\includegraphics[width=70.0mm]{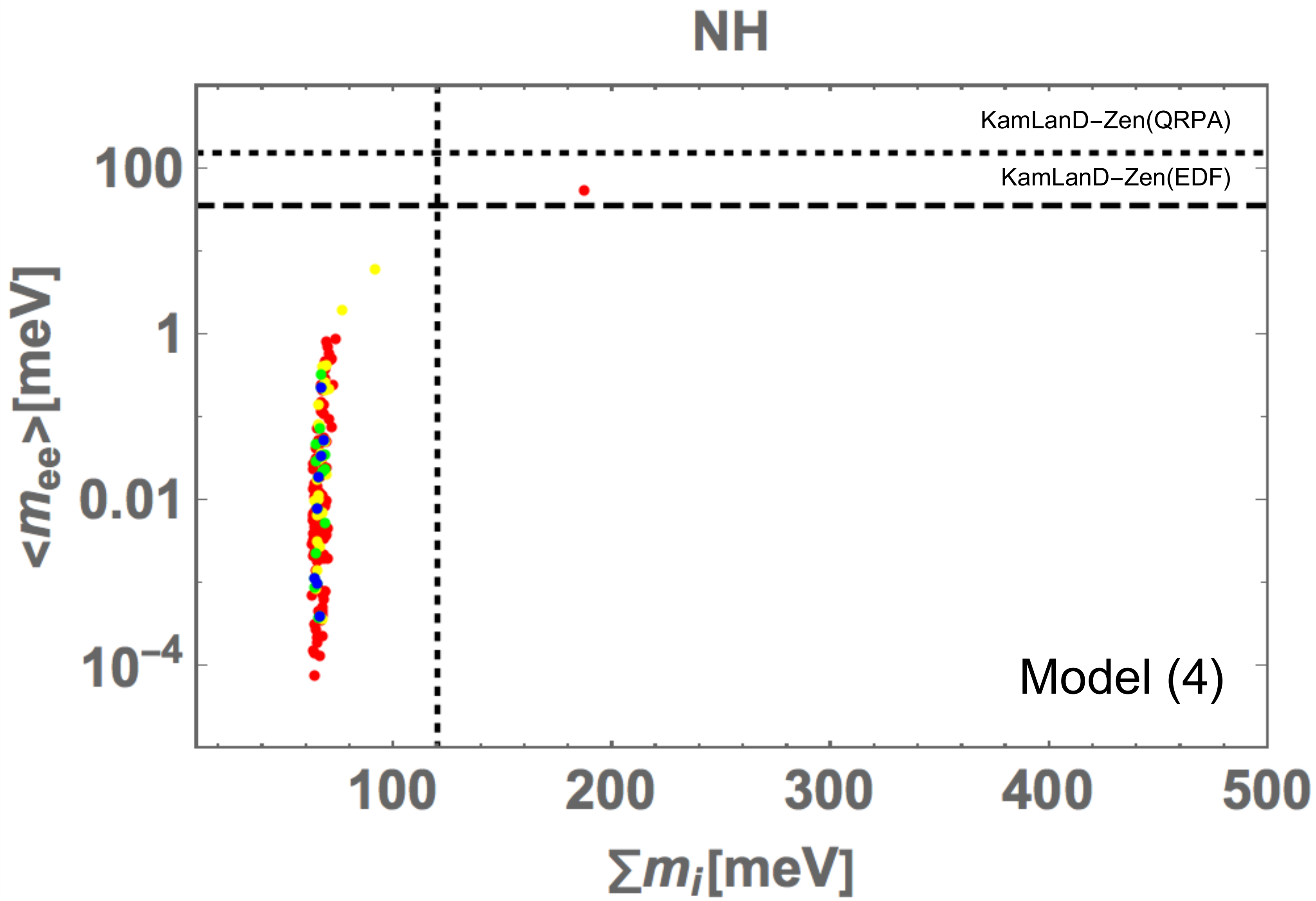} 
\includegraphics[width=70.0mm]{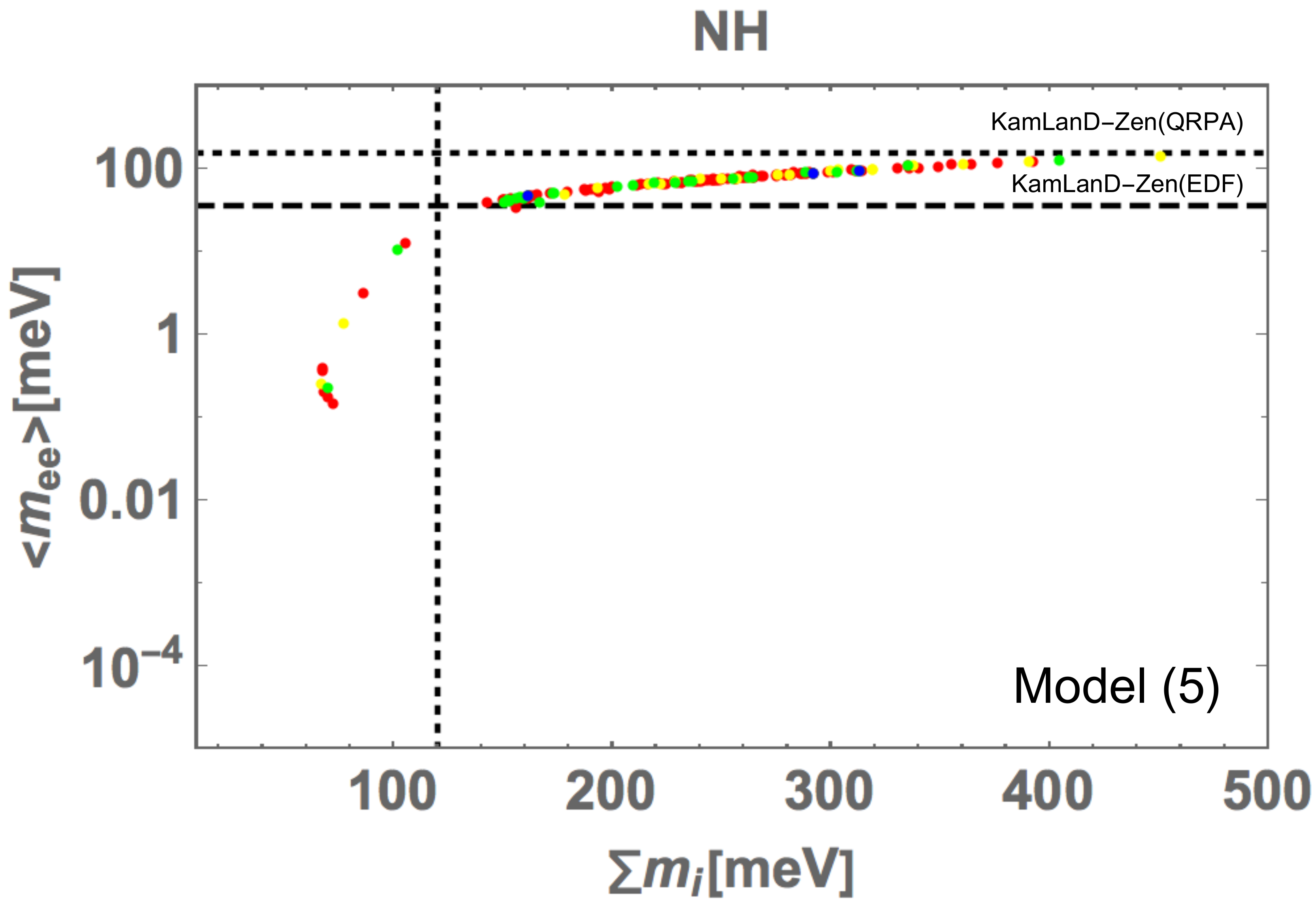} \quad
\includegraphics[width=70.0mm]{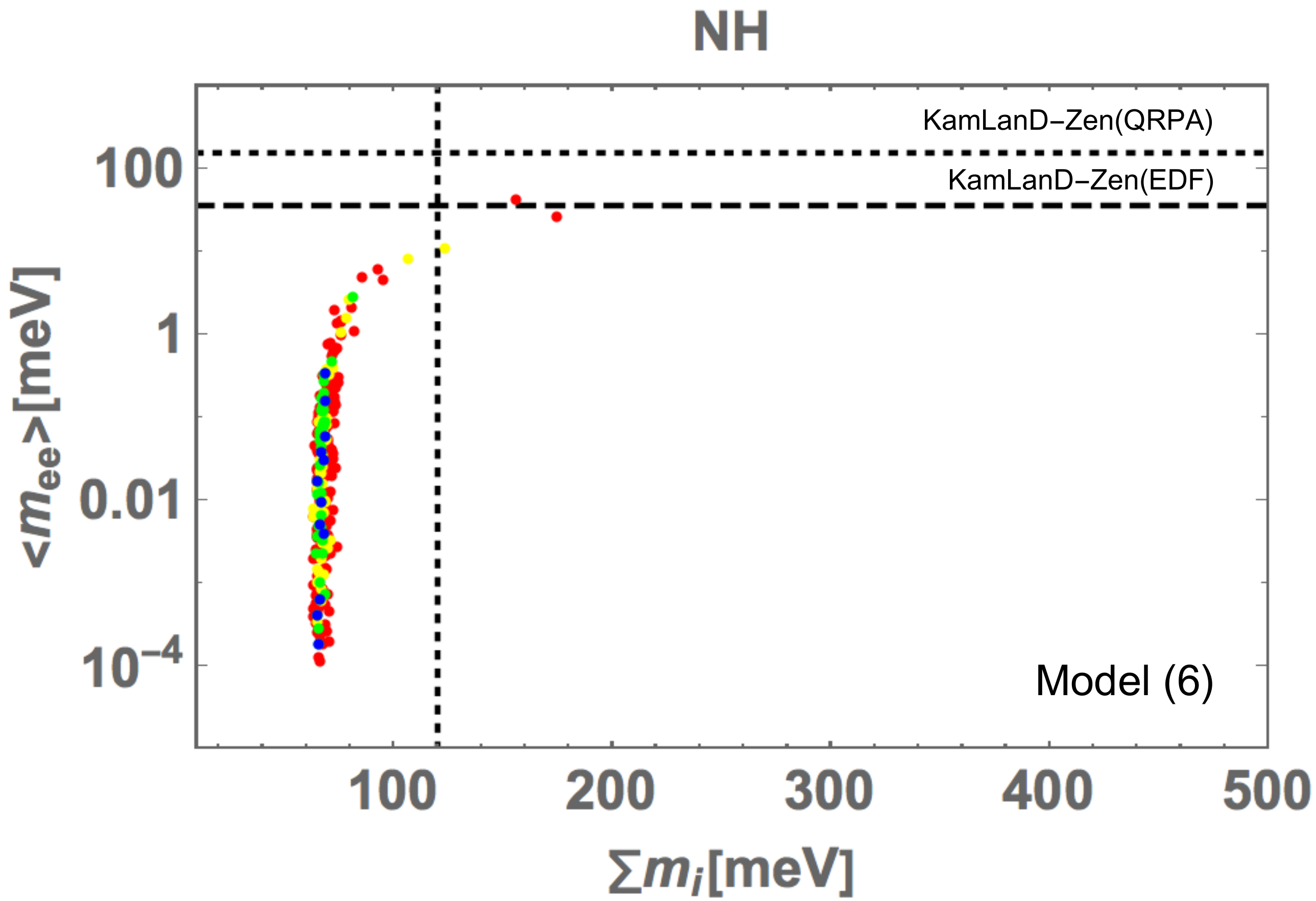} 
\caption{The correlations between $\sum m_i$ and $\langle m_{ee} \rangle$ in NH case for each model. In the plots, the horizontal lines are constraints from KamLAND-Zen where dashed one is the strongest constraint with energy-density functional (EDF) theory for nuclear matrix element and the dotted one is the weakest constraint with quasiparticle random-phase approximation (QRPA). The vertical dashed line corresponds to $\sum m_i = 120$ meV that is upper bound obtained from cosmology.}
  \label{fig:massesNH}
\end{center}\end{figure}

\begin{figure}[tb]
\begin{center}
\includegraphics[width=70.0mm]{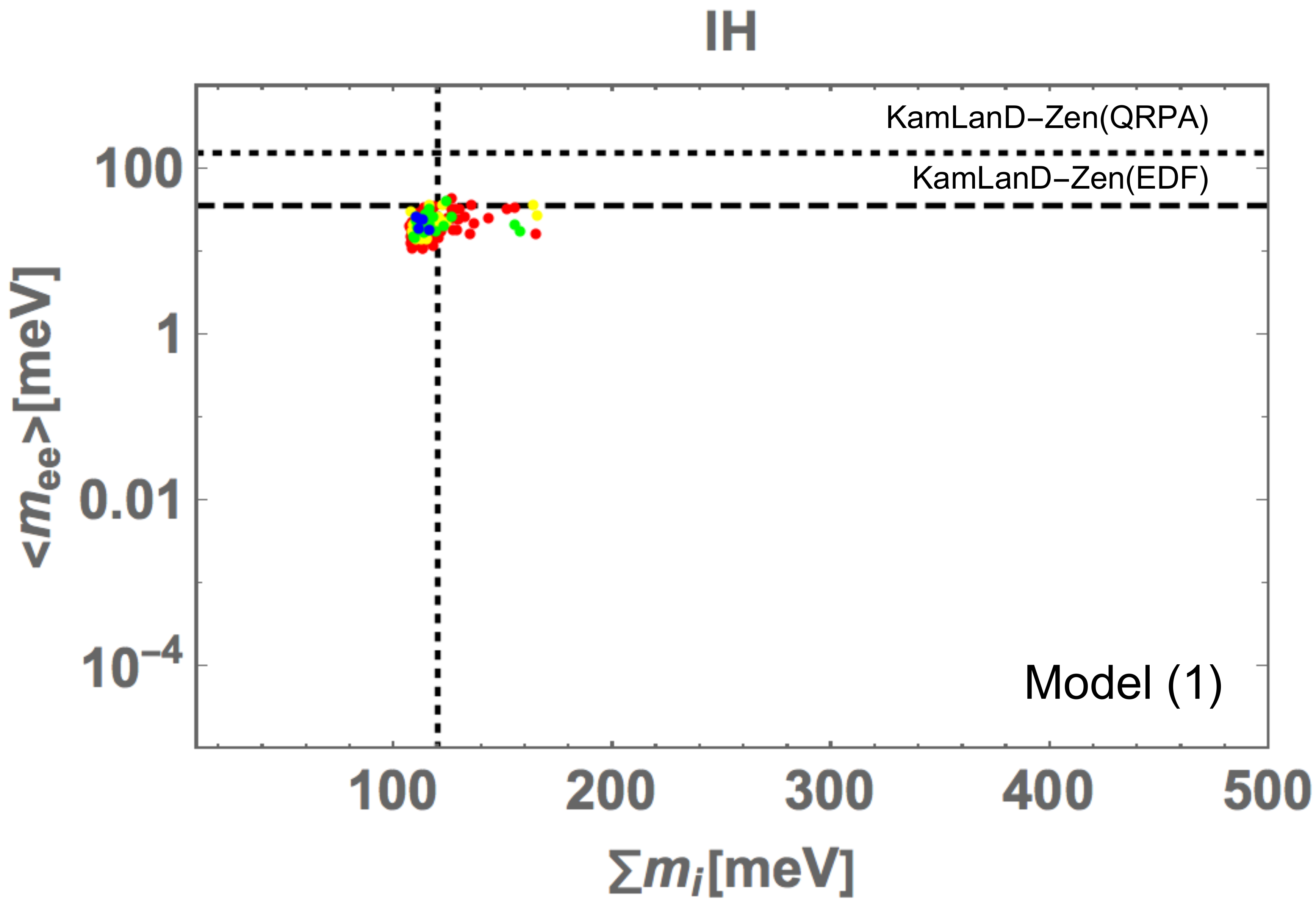} \quad
\includegraphics[width=70.0mm]{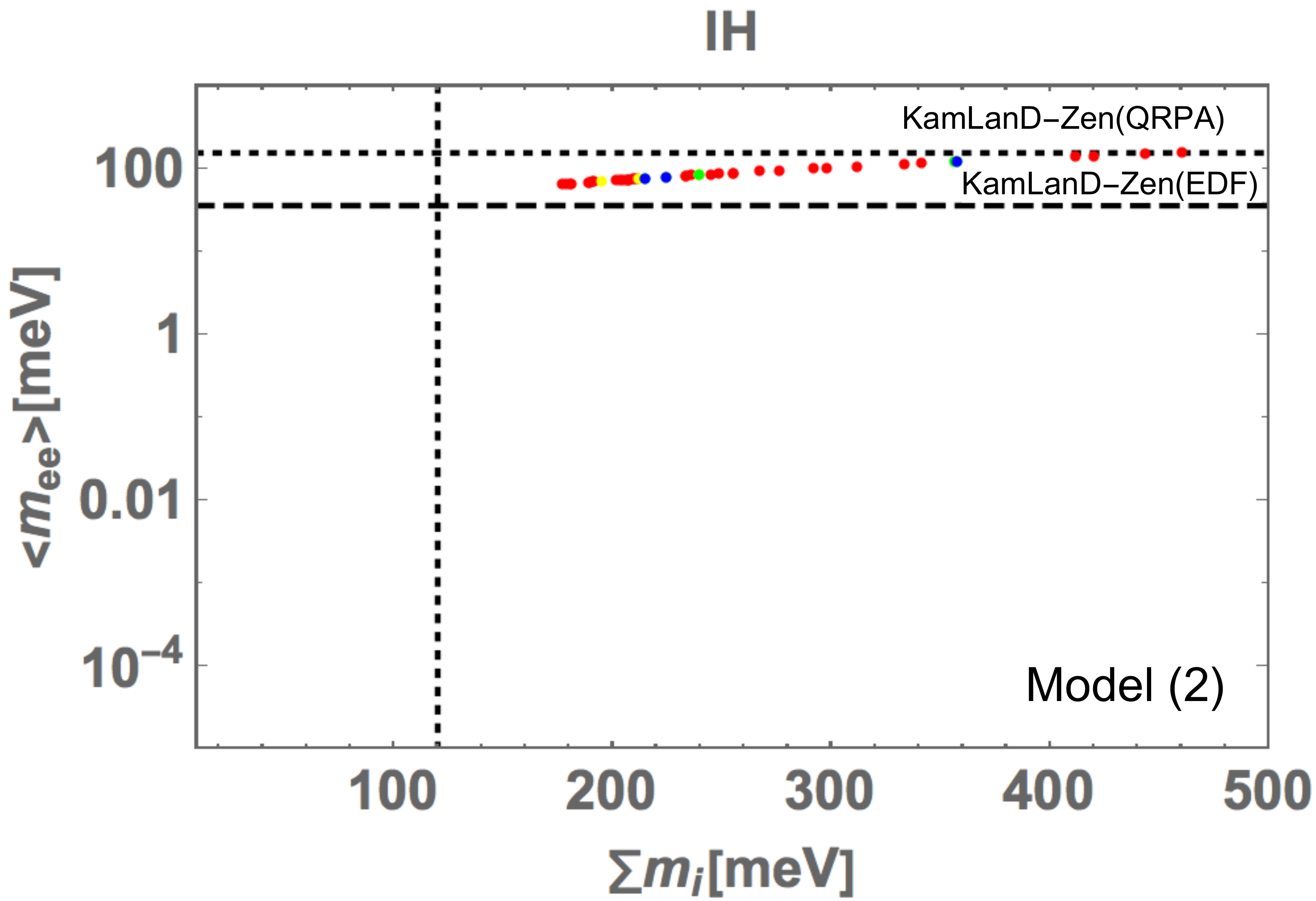} 
\includegraphics[width=70.0mm]{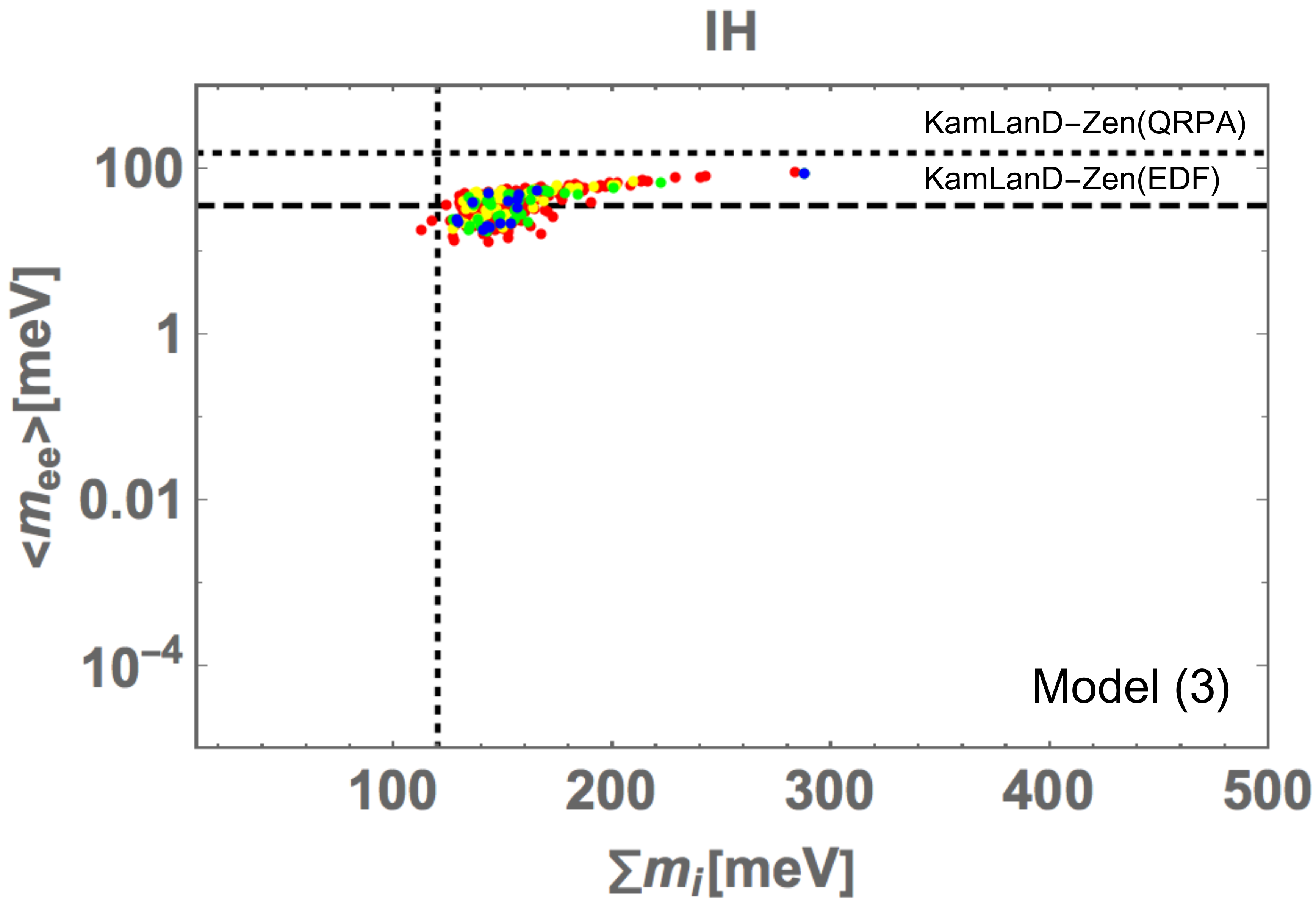} \quad
\includegraphics[width=70.0mm]{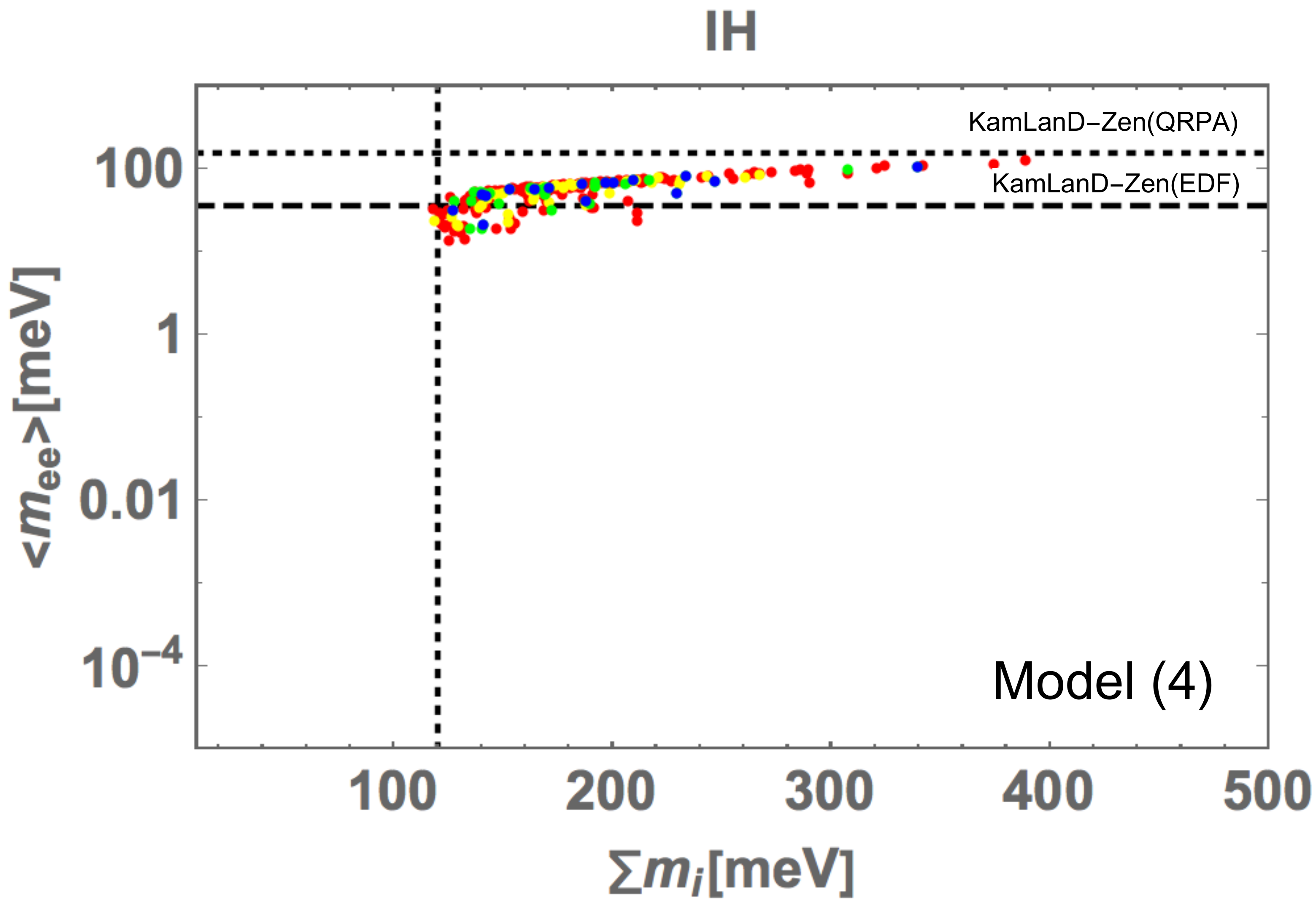} 
\includegraphics[width=70.0mm]{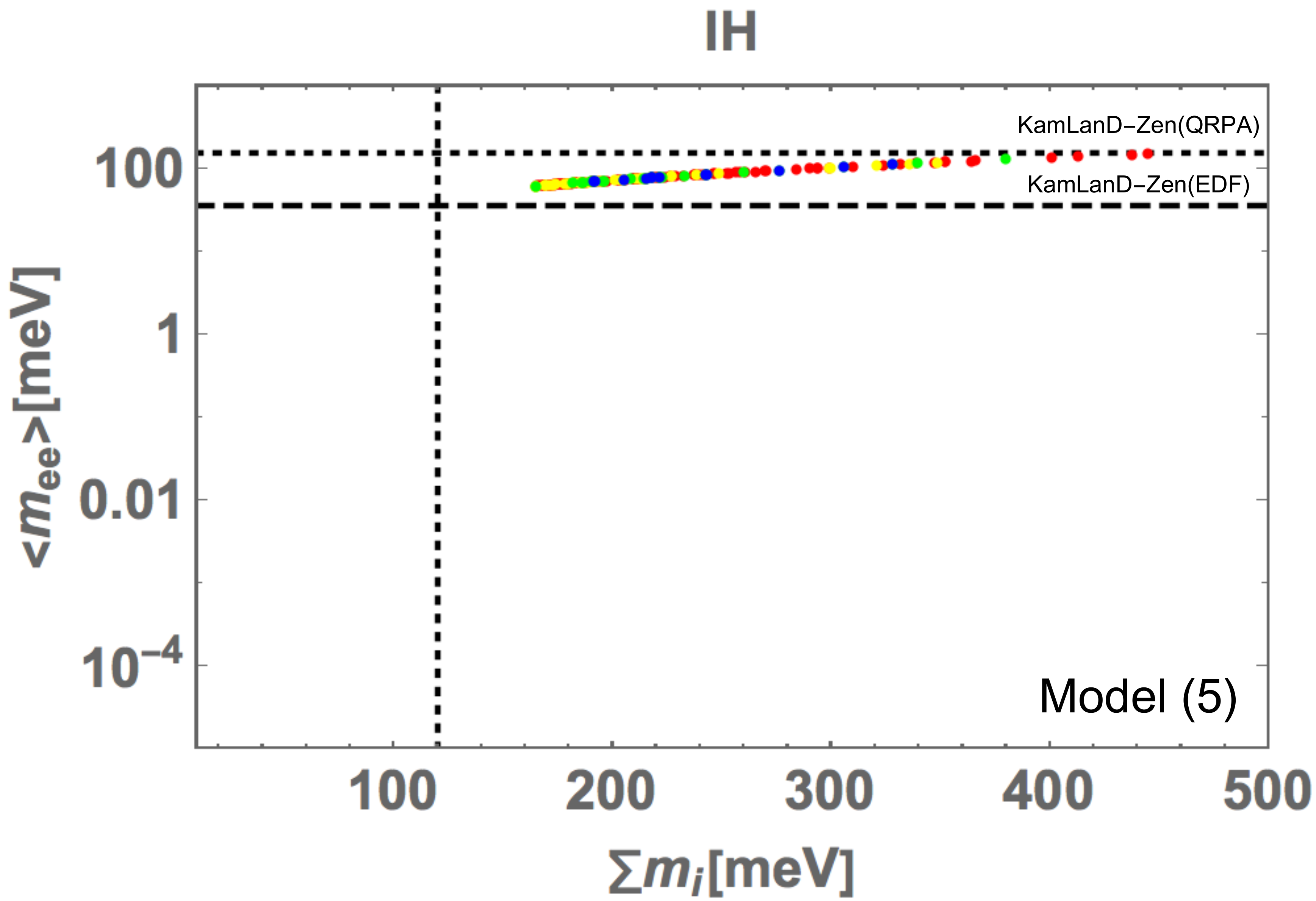} \quad
\caption{The correlations between $\sum m_i$ and $\langle m_{ee} \rangle$ in IH case for each model. The horizontal and vertical lines are the same as Fig.~\ref{fig:massesNH}.}
  \label{fig:massesIH}
\end{center}\end{figure}

In Fig.~\ref{fig:massesNH}, we show the correlation between $\sum m_i$ and $\langle m_{ee} \rangle$ in NH case for each model.
In the plots, the horizontal lines are constraints from KamLAND-Zen~\cite{KamLAND-Zen:2022tow} where dashed one is the strongest constraint with energy-density functional (EDF) theory for nuclear matrix element 
and the dotted one is the weakest constraint with quasiparticle random-phase approximation (QRPA).
In addition, the vertical dashed line corresponds to $\sum m_i = 120$ meV that is upper bound obtained from cosmology~\cite{Planck:2018vyg}.
For models except for (5), $\sum m_i$ tends to be around $[60, 80]$ meV where some points give larger value; in particular some amount of points in model (2) provide $\sum m_i \in [140, 210]$ meV.
On the other hand model (5) provides more points in larger $\sum m_i$ region up to 400 meV.
Also $\langle m_{ee} \rangle$ value tends to be small as $\langle m_{ee} \rangle \lesssim$ 1 meV except for models (2) and (5). 
Some points in model (2) and many points in model (5) provide $\langle m_{ee} \rangle = \mathcal{O}(10) $ meV to $\mathcal{O}(100)$ meV.
Note that in model (5) we do not have points satisfying $\sum m_i < 120$ meV and $\sigma < 1$. 
Also many points in model (5) and few points in other models are excluded by the strongest constraints on $\langle m_{ee} \rangle$ by KamLnd-Zen (allowed by the weakest constrains).
In Fig.~\ref{fig:massesIH}, we show the correlation between $\sum m_i$ and $\langle m_{ee} \rangle$ in IH case for each model.
For model (1), $\sum m_i$ tends to be around $[100, 170]$.
Models (3)[(4)] provide $\sum m_i \in [110, 300[400]]$ meV.
Model (2) and (5) tend to provide points in larger $\sum m_i$ region as [160, 460] meV.
On the other hand $\langle m_{ee} \rangle$ value tends to be $\mathcal{O}(10)$ meV to $\mathcal{O}(100)$ meV for models (1), (3) and (4), and $\mathcal{O}(100)$ meV for models (2) and (5).
Except for model (1) the most of the points are disfavored by cosmological constraint $\sum m_i < 120$ meV, and 
also many allowed points are excluded by the strongest constraints on $\langle m_{ee} \rangle$ by KamLnd-Zen while they are allowed by the weakest constrains;
in other words these cases are promising to be tested in near future experiments searching for neutrinoless double beta decay.

\begin{figure}[tb]
\begin{center}
\includegraphics[width=70.0mm]{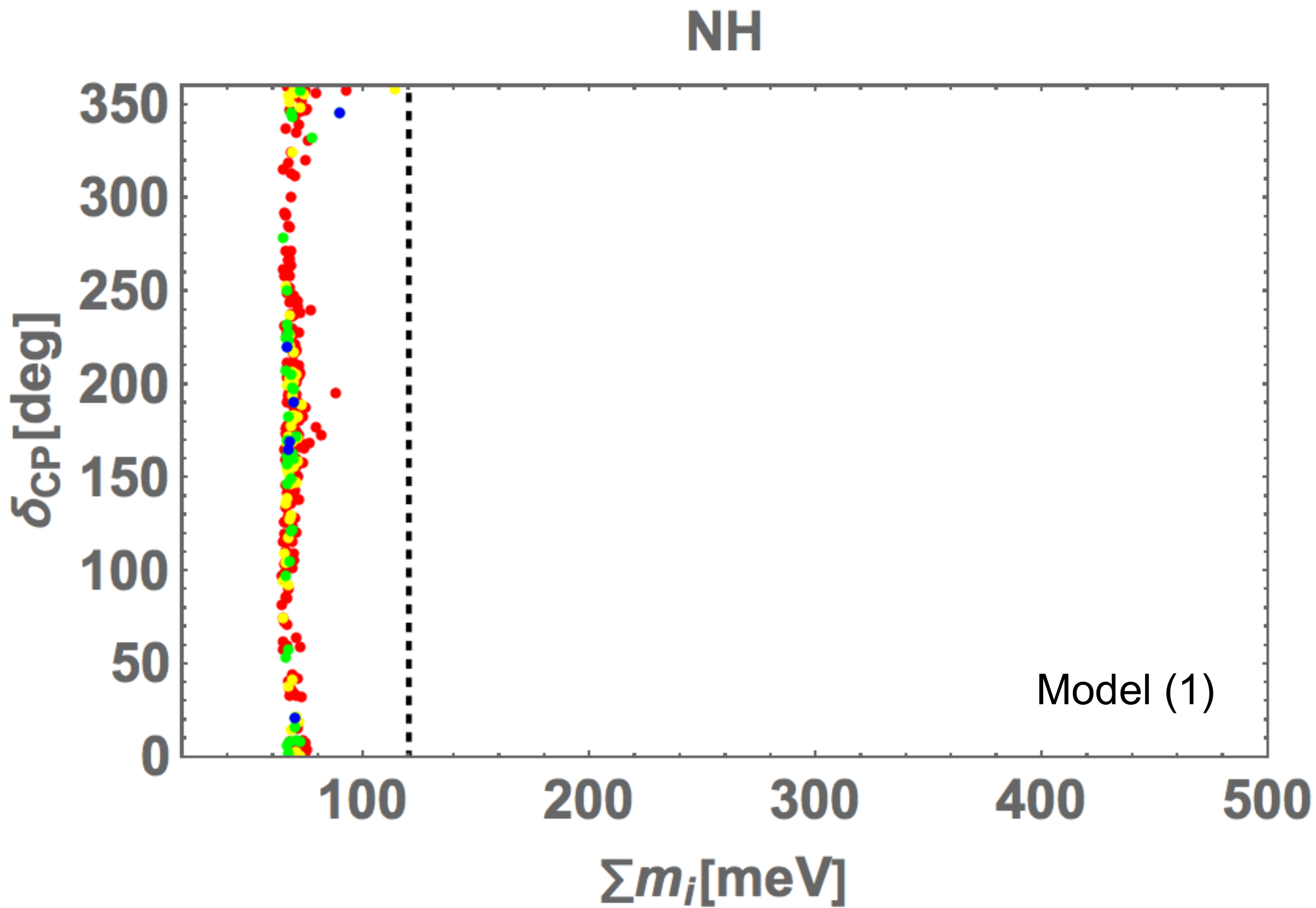} \quad
\includegraphics[width=70.0mm]{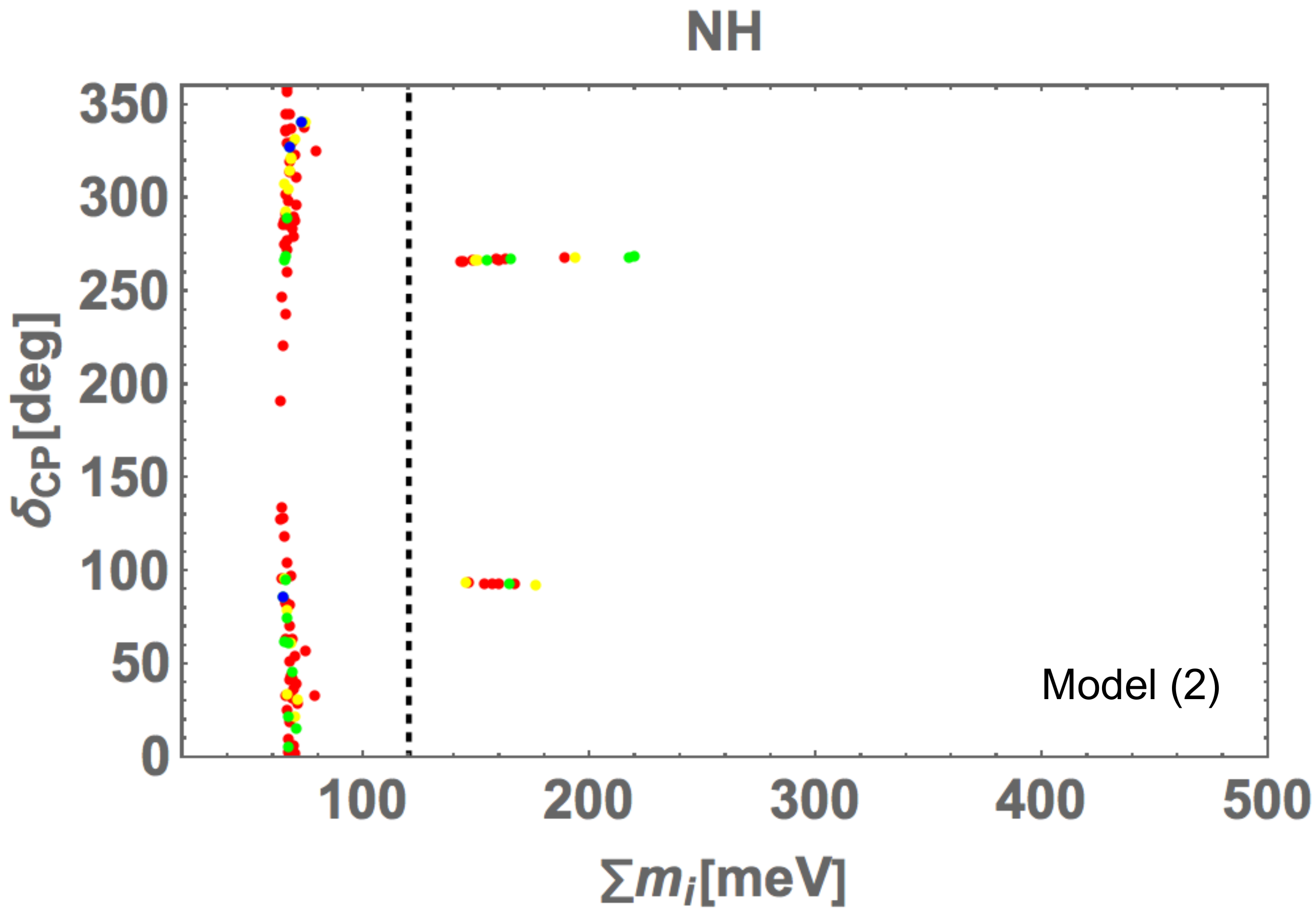} 
\includegraphics[width=70.0mm]{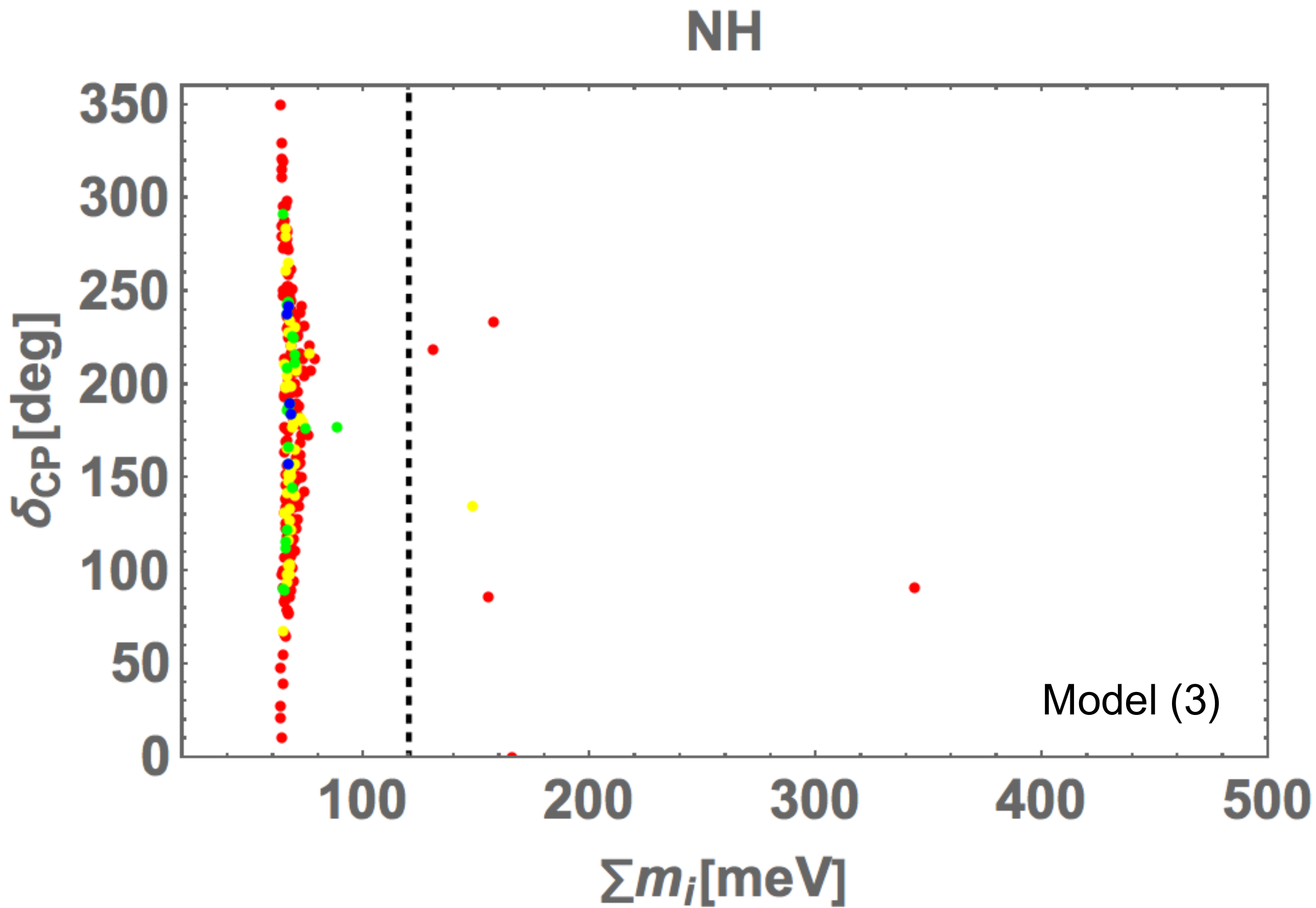} \quad
\includegraphics[width=70.0mm]{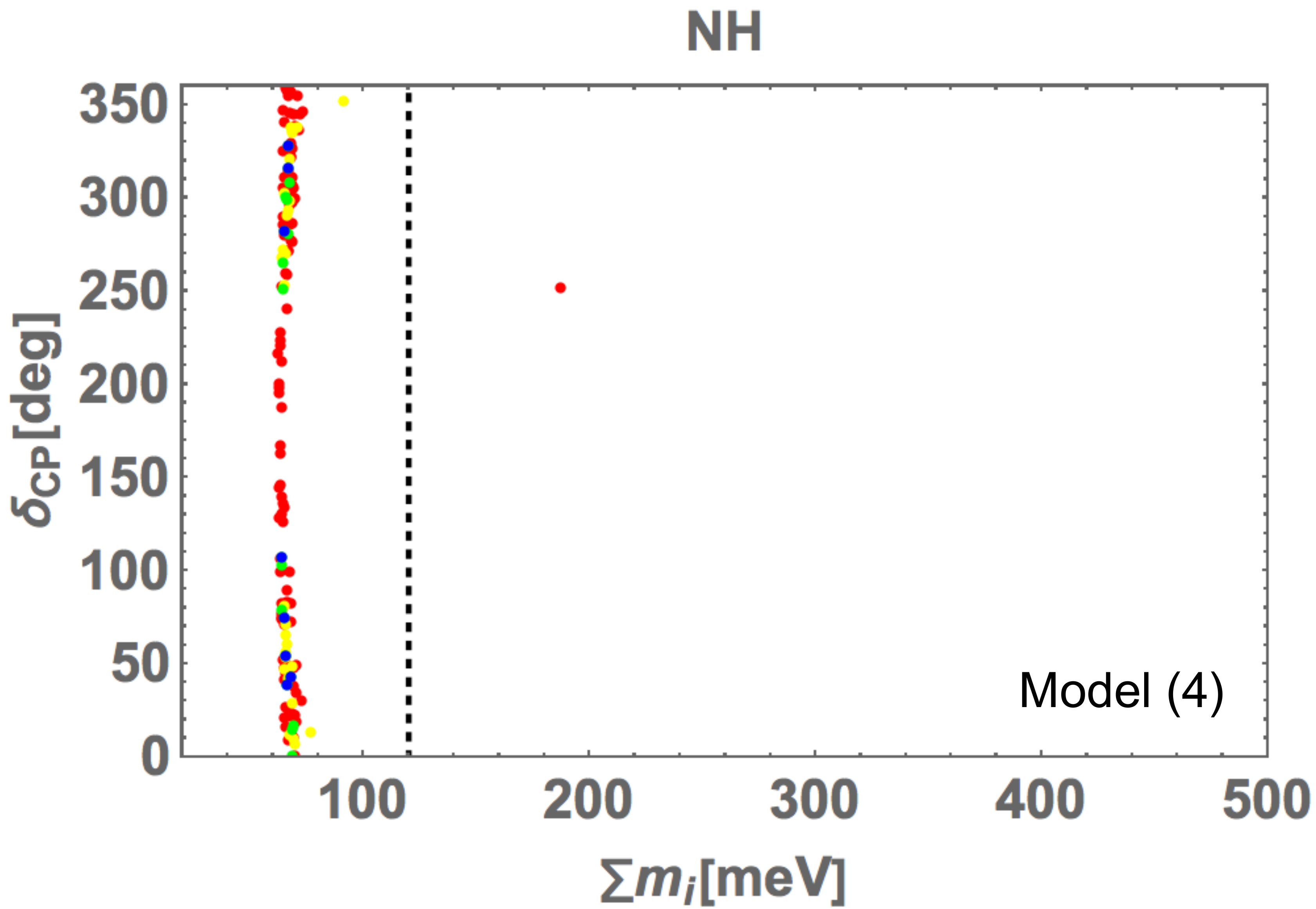} 
\includegraphics[width=70.0mm]{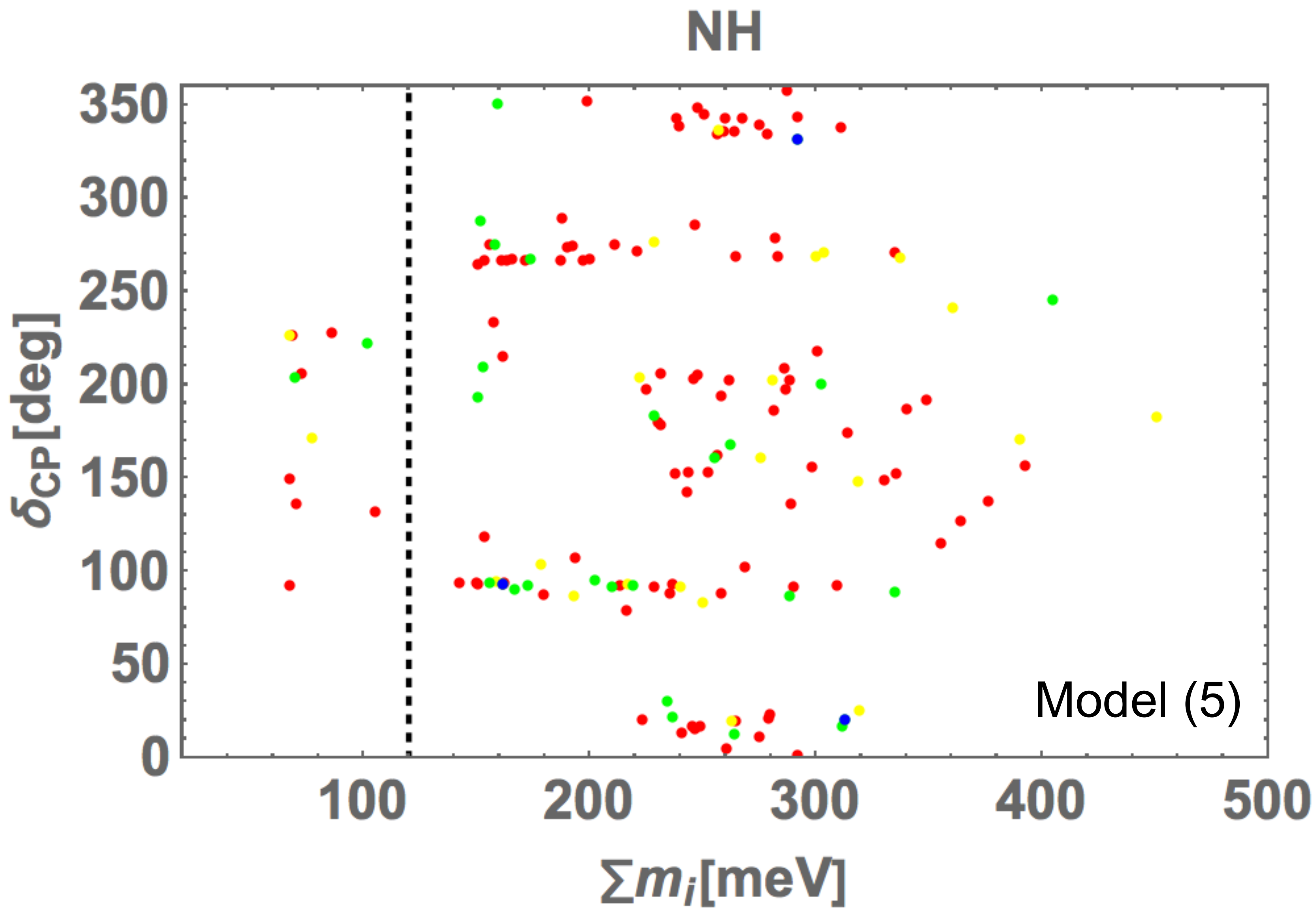} \quad
\includegraphics[width=70.0mm]{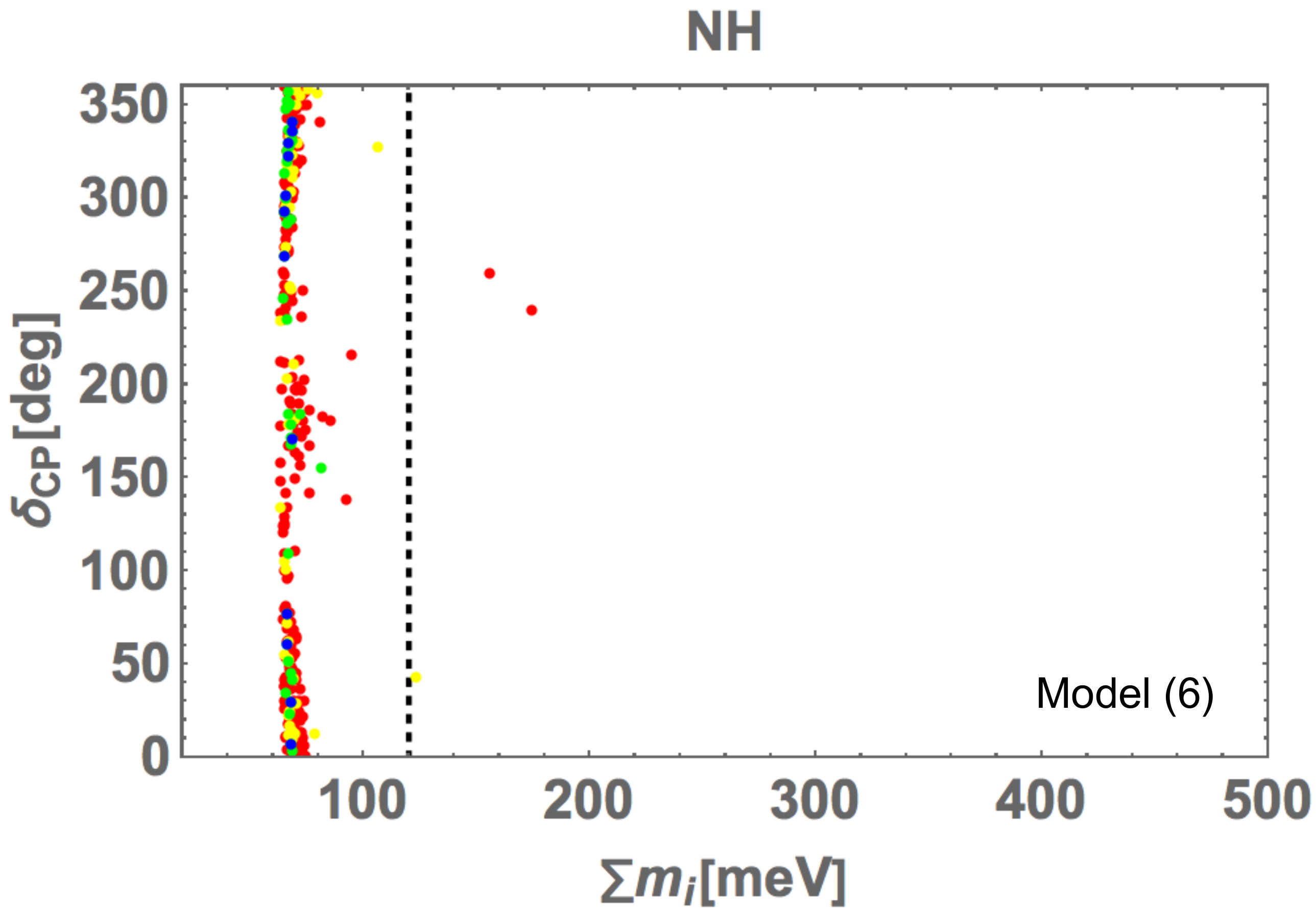} 
\caption{The correlations between $\sum m_i$ and $\delta_{CP}$ in NH case for each model. The vertical line is the same as Fig.~\ref{fig:massesNH}.}
  \label{fig:massphaseNH}
\end{center}\end{figure}

\begin{figure}[tb]
\begin{center}
\includegraphics[width=70.0mm]{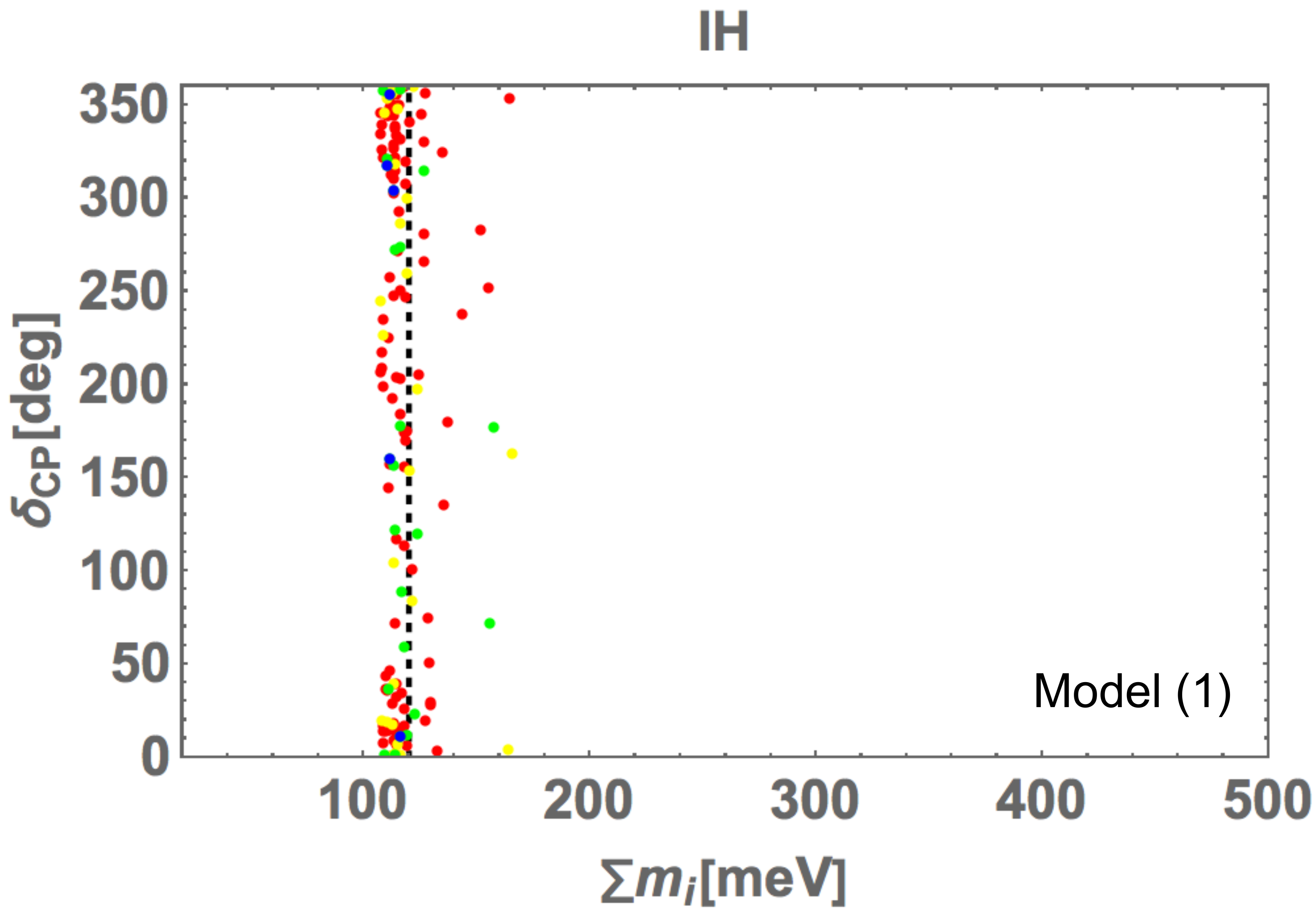} \quad
\includegraphics[width=70.0mm]{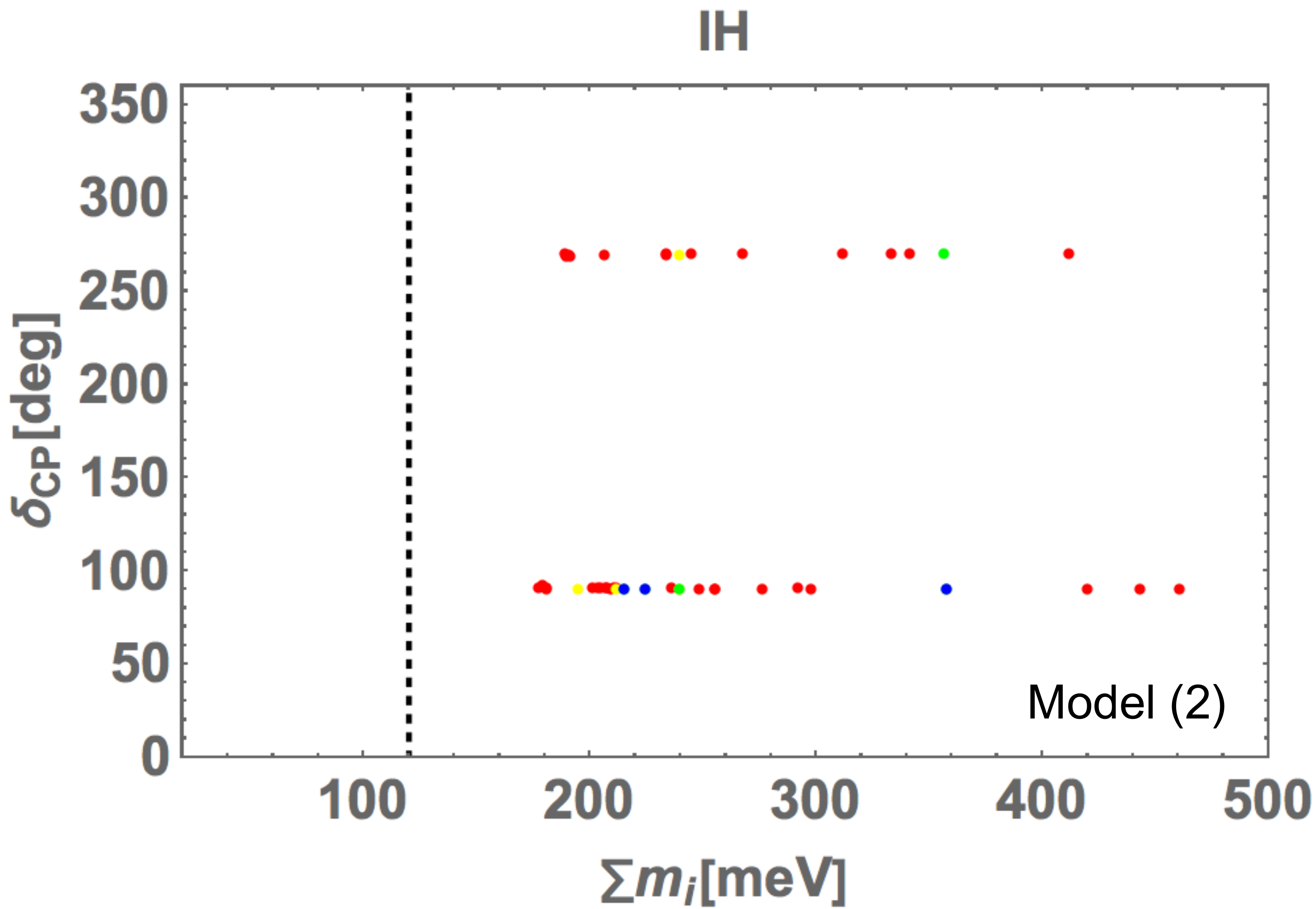} 
\includegraphics[width=70.0mm]{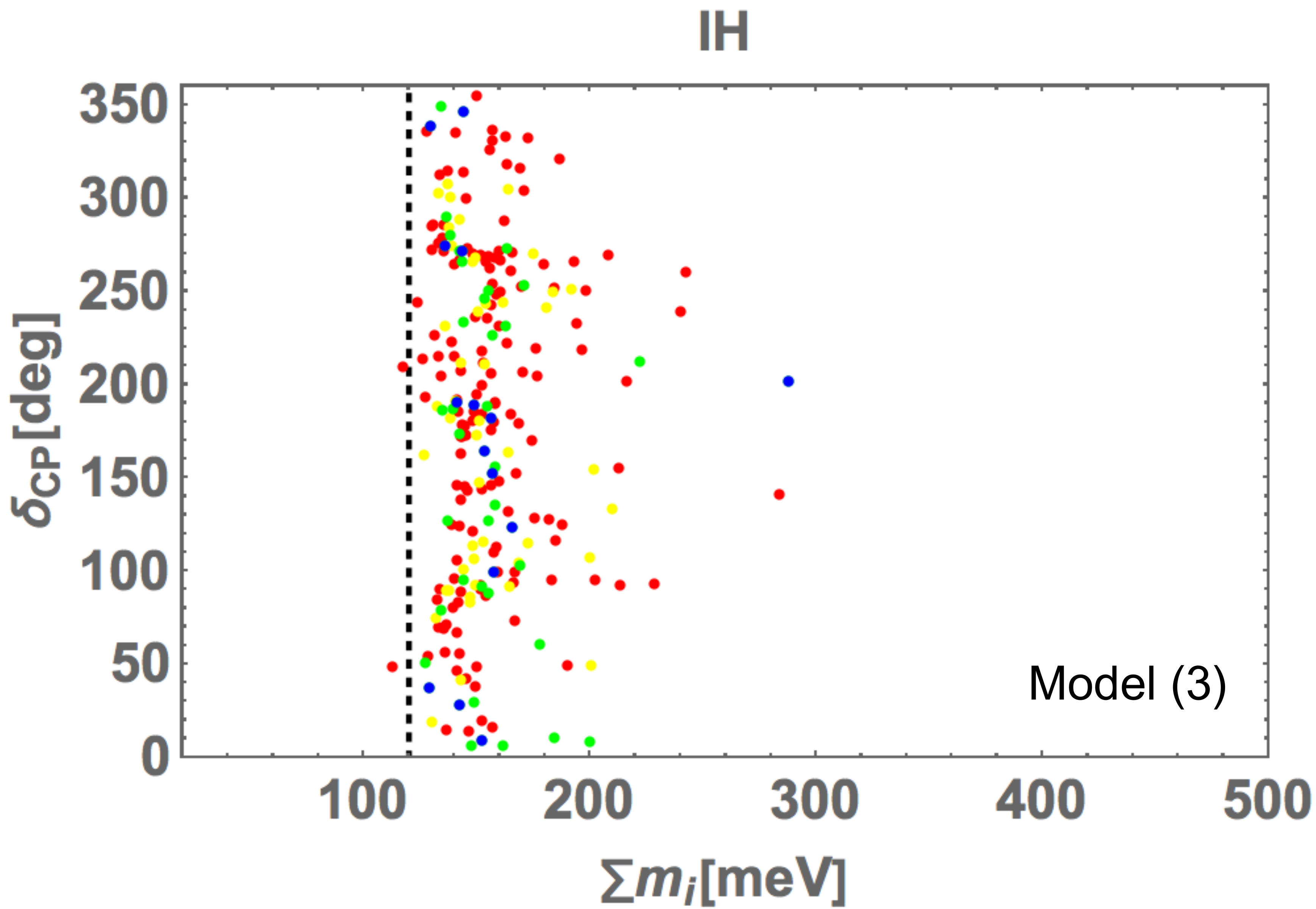} \quad
\includegraphics[width=70.0mm]{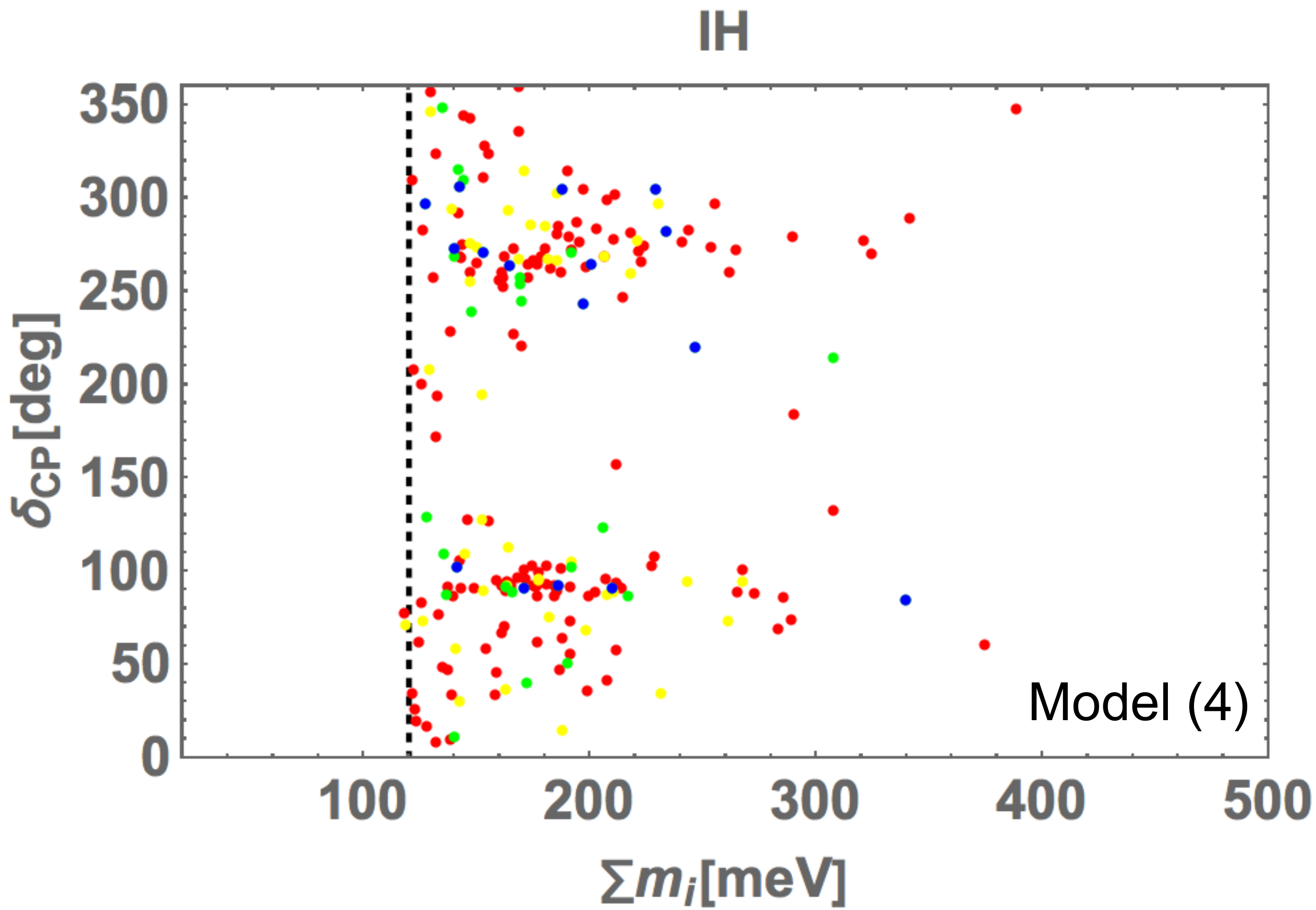} 
\includegraphics[width=70.0mm]{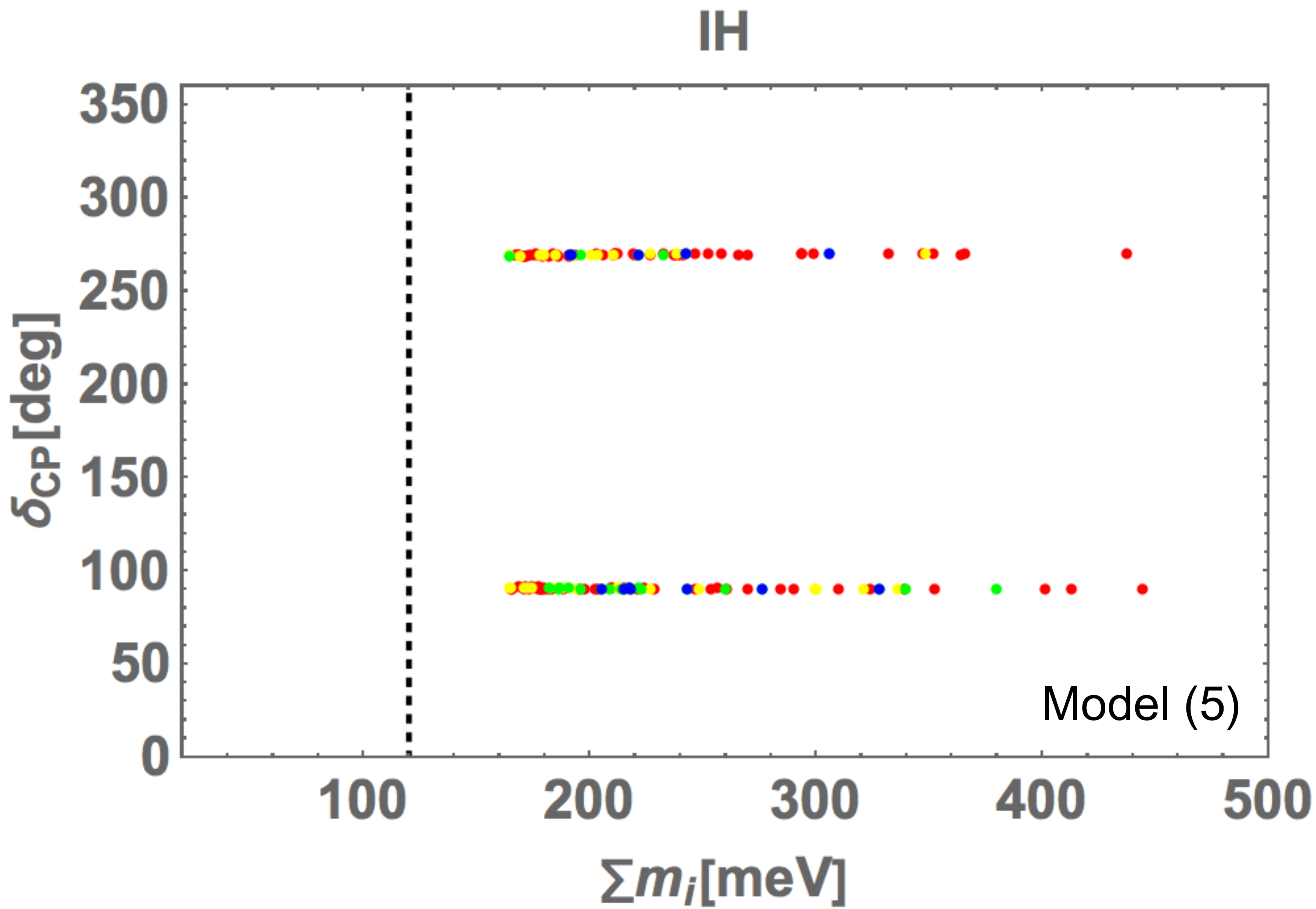}
\caption{The correlations between $\sum m_i$ and $\delta_{CP}$ in IH case for each model. The vertical line is the same as Fig.~\ref{fig:massesNH}.}
  \label{fig:massphaseIH}
\end{center}\end{figure}

In Figs.~\ref{fig:massphaseNH} and \ref{fig:massphaseIH} we show correlation between $\sum m_i$ and $\delta_{CP}$ for NH and IH cases respectively.
For NH case, we don't find large difference among models except for model (5) that gives points for $\sum m_i < 120$ meV when $90$[deg] $\lesssim \delta_{CP} \lesssim 230$[deg].
For IH case, we have more different patterns among models where models (2) and (5) provide similar pattern with $\delta_{CP}$ being localized around $90$[deg] and $270$[deg].
The model (1) almost satisfies $\sum m_i\le$ 120 meV.

We also comment on the hierarchical structure of the Yukawa coupling $y_i$ for the allowed parameter sets.
For NH case, we find 
\begin{itemize}
\item Model (1) : Hierarchy among $y_i$ is not fixed.
\item  Model (2) : $|y_1| \ll \{|y_2|, |y_3| \}$.
\item Model (3) : Hierarchy among $|y_i|$ is not fixed.
\item Model (4) : Hierarchy among $|y_i|$ is not fixed, but we have more points for $|y_1| \ll |y_3| < |y_2|$.
\item Model (5) : Most points indicate $|y_1| \sim |y_2| \sim |y_3|$ and few points correspond to $|y_1| \ll \{|y_2|, |y_3| \}$.
\item Model (6) : Hierarchy among $|y_i|$ is not fixed, but we have more points for $|y_1| \ll |y_3| < |y_2|$.
\end{itemize}
For IH case we find
\begin{itemize}
\item Model (1) : $|y_1| \sim |y_2|$ and $|y_3|$ can be $|y_3| /|y_1| \ll 1$ to $|y_3|/|y_1| \sim 90$.
\item  Model (2) :$|y_1| \sim |y_2|$ and $|y_3|$ can be $|y_3| /|y_1| \ll 1$ to $|y_3|/|y_1| \sim 20$.
\item Model (3) : $|y_1| \sim |y_2| \sim |y_3|$.
\item Model (4) : $|y_1| \sim |y_3| \lesssim |y_2|$.
\item Model (5) : $|y_2| < |y_1|< |y_3|$.
\end{itemize}
These structures determine strength of Yukawa interactions $\bar L \eta N_R$ and it could be tested from decay pattern of new particles $\eta$ and $\psi_i$ at collider experiments.
Exploring such a possibility is beyond the scope of this analysis.

\subsection{Implications to phenomenology}

Here we discuss some phenomenological implications in the model.
We have $Z'$ boson in the model that interacts with the SM fermions $f$ with lepton flavor dependent way.
The decay width for $Z' \to \bar f f$ is given by 
\begin{equation}
\Gamma (Z' \to \bar f f ) = N_c \frac{g_X^2 Q_X^2}{12 \pi} \left( 1 + \frac{2 m_f^2}{m_{Z'}^2} \right) \sqrt{1 - \frac{4 m_f^2}{m_{Z'}^2}},
\end{equation}
where $m_f$ and $Q_X$ are a mass and a $U(1)_X$ charge of fermion $f$, and $N_c$ is the number of color.
The $Z'$ boson can be produced at the LHC via $q \bar q \to Z'$ the same as $Z'$ in $U(1)_{B-L}$ case. 
We then consider $m_{Z'}$ is sufficiently larger than masses of the SM fermions; typically $m_{Z'} > 5$ TeV to avoid current constraints from the LHC data~\cite{ATLAS:2019erb,CMS:2021ctt}. 
Assuming other new particles, $\{\psi_i, \eta_{I/R}, \eta^\pm \}$, are heavier than $m_{Z'}/2$, the branching ratios (BRs) of the $Z'$ decay are given by 
\begin{align}
& BR(Z' \to \bar q q) : BR(Z' \to \bar \nu \nu): BR(Z' \to e^+ e^-) : BR(Z' \to \mu^+ \mu^-) : BR(Z' \to \tau^+ \tau^-)  \nonumber \\
& = 1 : \frac32 (x^2 +y^2 +z^2) : 3x^2 : 3y^2 : 3z^2,
\end{align}
where three generations of neutrinos are summed for BR of $Z' \to \bar \nu \nu$.
Therefore we can distinguish the models if we observe $Z'$ decaying into leptons and investigate ratio of BRs.
The ratios of BRs for leptonic modes are summarized in Table~\ref{tab:2}.

A DM candidate in the model is the lightest neutral particle that is $Z_2$ odd; the lightest component of $\psi_i$ or $\eta_{I/R}$.
The physics of DM candidates is mostly the same as the original scotogenic model.
The difference is that we have $Z'$ interaction in the models. 
We thus have more freedom to fit relic density of DM in the models tuning $Z'$ mass and new gauge coupling $g_X$.
Here we leave detailed analysis in our future work.

\begin{table}[t!]
\begin{tabular}{|c||c|c|c|c|c|}\hline\hline  
& ~$(\chi^2)^{NH}_{\rm min}$~ & ~$(\chi^2)^{IH}_{\rm min}$~ & ~$\sum m_i$/meV (NH  [IH])~ & ~$\langle m_{ee} \rangle$/meV (NH  [IH])~ & ~$BR_{ee} : BR_{\mu \mu} : BR_{\tau \tau}$~ \\\hline\hline 
model (1) & 2.13 & 0.539 & $113 \, [113]$ & $0.00648 \, [24.1]$ & 1 : 1 : 9  \\ \hline
model (2) & 3.46 & 2.76 & $67.5 \, [357]$ & $0.0332 \, [122]$ & 1 : 1 : 9  \\ \hline
model (3) & 2.72 & 0.228 & $66.5 \, [288]$ & $0.00487 \, [89.0]$  & 9 : 1 : 1  \\ \hline
model (4) & 2.82  & 0.208 & $66.1 \, [210]$ & $0.000387 \, [73.3]$ & 9 : 1 : 1  \\ \hline
model (5) & 3.13 & 0.830 & $292 \, [215]$ & $89.2 \, [77.1]$  & 1 : 9 : 1  \\ \hline
model (6) & 1.06 & 79.6  & $65.2 \, [--]$ & $0.0171 \, [--]$  & 1 : 9 : 1  \\ \hline
\end{tabular}
\caption{ Summary of the benchmarks giving minimal $\chi^2$ showing $\sum m_i$ and $\langle m_{ee} \rangle$ for NH and IH cases, and BRs of $Z' \to \bar \ell \ell$,  in each model. }\label{tab:2}
\end{table}

\section{Summary and discussion}

We have discussed scotogenic models with a general lepton flavor dependent $U(1)_X$ gauge symmetry where $X = B - x L_e - y L_\mu - z L_\tau$ with $x+y+z =3$ to cancel anomalies.
Flavor structures of Yukawa couplings and the Majorana mass matrix of the SM singlet fermions are restricted by the $U(1)_X$ charge assignments.
In the analysis of this study, we have chosen the $U(1)_X$ charges and field contents as follows; (1) Yukawa couplings of the terms $\bar L_L \eta N_R$ and $\bar L_L H e_R$ are diagonal, 
(2) The Majorana mass matrix of $N_R$ has two-zero textures, (3) only one singlet scalar $\varphi$ has to break $U(1)_X$ for minimality, (4) all three generations of leptons have non-zero $U(1)_X$ charges.
Then we have found the six models that satisfy the above criteria.

We have carried out numerical analysis of neutrino mass matrix which is induced at one-loop level for each texture, and searched for allowed parameter sets which can accommodate neutrino data.
Some specific correlations are found for CP phases $\alpha_{21}$ and $\delta_{CP}$ due to the two-zero texture of Majorana mass matrix of $N_R$.
We have also shown $\sum m_i$ and $\langle m_{ee} \rangle$ where characteristic predicted region are found depending on the models.
It is found that the most points in the NH case of model (5) and the IH cases except for model (1) are disfavored by cosmological constraint on sum of neutrino masses.
Also many allowed points in the IH cases and the NH case of model (5) are excluded by the strongest constraints on $\langle m_{ee} \rangle$ by KamLnd-Zen (allowed by the weakest constrains);
in other words these cases are promising to be tested in near future experiments searching for neutrinoless double beta decay. 

These models have $Z'$ boson that provide specific ratios of BRs for decay $Z' \to \bar f f$ due to our charge assignment.
In particular, the ratios of BRs for leptonic modes show a specific pattern for each model.
Thus exploring the BRs, we can test the models in addition to predictions in neutrino sector, which are summarized in Table~\ref{tab:2}.
The DM in the models also interacts with $Z'$ and DM physics would be modified from original scotgenic model 
where we leave detailed analysis in future work.

\section*{Acknowledgments}
This research was supported by an appointment to the JRG Program at the APCTP through the Science and Technology Promotion Fund and Lottery Fund of the Korean Government. This was also supported by the Korean Local Governments - Gyeongsangbuk-do Province and Pohang City (H.O.). 
H. O. is sincerely grateful for KIAS and all the members.
The work was also supported by the Fundamental Research Funds for the Central Universities (T.~N.).


\begin{thebibliography}{99}


\bibitem{Ma:2006km} 
  E.~Ma,
  Phys.\ Rev.\ D {\bf 73}, 077301 (2006)
  [hep-ph/0601225].
 
  
\bibitem{Kajiyama:2013zla} 
  Y.~Kajiyama, H.~Okada and K.~Yagyu,
  Nucl.\ Phys.\ B {\bf 874}, 198 (2013)
  [arXiv:1303.3463 [hep-ph]].

\bibitem{Krauss:2002px} 
  L.~M.~Krauss, S.~Nasri and M.~Trodden,
  Phys.\ Rev.\ D {\bf 67}, 085002 (2003)
  [hep-ph/0210389].



\bibitem{Aoki:2008av} 
  M.~Aoki, S.~Kanemura and O.~Seto,
  Phys.\ Rev.\ Lett.\  {\bf 102}, 051805 (2009)
  [arXiv:0807.0361 [hep-ph]].
  
\bibitem{Gustafsson:2012vj} 
  M.~Gustafsson, J.~M.~No and M.~A.~Rivera,
  Phys.\ Rev.\ Lett.\  {\bf 110}, no. 21, 211802 (2013)
  Erratum: [Phys.\ Rev.\ Lett.\  {\bf 112}, no. 25, 259902 (2014)]
  doi:10.1103/PhysRevLett.110.211802, 10.1103/PhysRevLett.112.259902
  [arXiv:1212.4806 [hep-ph]].
  
  
\bibitem{Fritzsch:2011qv}
H.~Fritzsch, Z.~z.~Xing and S.~Zhou,
JHEP \textbf{09} (2011), 083
doi:10.1007/JHEP09(2011)083
[arXiv:1108.4534 [hep-ph]].


\bibitem{Branco:1988ex}
G.~C.~Branco, W.~Grimus and L.~Lavoura,
Nucl. Phys. B \textbf{312} (1989), 492-508
doi:10.1016/0550-3213(89)90304-0

\bibitem{Choubey:2004hn}
S.~Choubey and W.~Rodejohann,
Eur. Phys. J. C \textbf{40} (2005), 259-268
doi:10.1140/epjc/s2005-02133-1
[arXiv:hep-ph/0411190 [hep-ph]].

  
\bibitem{Araki:2012ip}
T.~Araki, J.~Heeck and J.~Kubo,
JHEP \textbf{07} (2012), 083
doi:10.1007/JHEP07(2012)083
[arXiv:1203.4951 [hep-ph]].


\bibitem{Baek:2015mna}
S.~Baek, H.~Okada and K.~Yagyu,
JHEP \textbf{04} (2015), 049
doi:10.1007/JHEP04(2015)049
[arXiv:1501.01530 [hep-ph]].

\bibitem{Crivellin:2015lwa}
A.~Crivellin, G.~D'Ambrosio and J.~Heeck,
Phys. Rev. D \textbf{91} (2015) no.7, 075006
doi:10.1103/PhysRevD.91.075006
[arXiv:1503.03477 [hep-ph]].

\bibitem{Plestid:2016esp}
R.~Plestid,
Phys. Rev. D \textbf{93} (2016) no.3, 035011
doi:10.1103/PhysRevD.93.035011
[arXiv:1602.06651 [hep-ph]].


  
\bibitem{Ko:2017quv}
P.~Ko, T.~Nomura and H.~Okada,
Phys. Lett. B \textbf{772} (2017), 547-552
doi:10.1016/j.physletb.2017.07.021
[arXiv:1701.05788 [hep-ph]].

  
  
\bibitem{Lee:2017ekw}
S.~Lee, T.~Nomura and H.~Okada,
Nucl. Phys. B \textbf{931} (2018), 179-191
doi:10.1016/j.nuclphysb.2018.04.010
[arXiv:1702.03733 [hep-ph]].

\bibitem{Asai:2017ryy}
K.~Asai, K.~Hamaguchi and N.~Nagata,
Eur. Phys. J. C \textbf{77} (2017) no.11, 763
doi:10.1140/epjc/s10052-017-5348-x
[arXiv:1705.00419 [hep-ph]].

  
\bibitem{Nomura:2017psk}
T.~Nomura and H.~Okada,
Phys. Rev. D \textbf{97} (2018) no.5, 055044
doi:10.1103/PhysRevD.97.055044
[arXiv:1707.06083 [hep-ph]].

  
  
\bibitem{Nomura:2018vfz}
T.~Nomura and H.~Okada,
Phys. Rev. D \textbf{97} (2018) no.9, 095023
doi:10.1103/PhysRevD.97.095023
[arXiv:1803.04795 [hep-ph]].



\bibitem{Nomura:2018cle}
T.~Nomura and H.~Okada,
Phys. Lett. B \textbf{783} (2018), 381-386
doi:10.1016/j.physletb.2018.07.011
[arXiv:1805.03942 [hep-ph]].



\bibitem{Ko:2019tts}
P.~Ko, T.~Nomura and C.~Yu,
JHEP \textbf{04} (2019), 102
doi:10.1007/JHEP04(2019)102
[arXiv:1902.06107 [hep-ph]].

\bibitem{Asai:2018ocx}
K.~Asai, K.~Hamaguchi, N.~Nagata, S.~Y.~Tseng and K.~Tsumura,
Phys. Rev. D \textbf{99} (2019) no.5, 055029
doi:10.1103/PhysRevD.99.055029
[arXiv:1811.07571 [hep-ph]].


\bibitem{Han:2019lux}
Z.~L.~Han and W.~Wang,
Eur. Phys. J. C \textbf{79} (2019) no.6, 522
doi:10.1140/epjc/s10052-019-7033-8
[arXiv:1901.07798 [hep-ph]].


\bibitem{Nomura:2019dhw}
T.~Nomura and K.~Yagyu,
JHEP \textbf{10} (2019), 105
doi:10.1007/JHEP10(2019)105
[arXiv:1905.11568 [hep-ph]].



\bibitem{Asai:2019ciz}
K.~Asai,
Eur. Phys. J. C \textbf{80} (2020) no.2, 76
doi:10.1140/epjc/s10052-020-7622-6
[arXiv:1907.04042 [hep-ph]].





\bibitem{Araki:2019rmw}
T.~Araki, K.~Asai, J.~Sato and T.~Shimomura,
Phys. Rev. D \textbf{100} (2019) no.9, 095012
doi:10.1103/PhysRevD.100.095012
[arXiv:1909.08827 [hep-ph]].


\bibitem{Han:2019diw}
Z.~L.~Han, R.~Ding, S.~J.~Lin and B.~Zhu,
Eur. Phys. J. C \textbf{79} (2019) no.12, 1007
doi:10.1140/epjc/s10052-019-7526-5
[arXiv:1908.07192 [hep-ph]].



\bibitem{Wang:2019byi}
W.~Wang and Z.~L.~Han,
Phys. Rev. D \textbf{101} (2020) no.11, 115040
doi:10.1103/PhysRevD.101.115040
[arXiv:1911.00819 [hep-ph]].


\bibitem{Chen:2020jvl}
C.~H.~Chen and T.~Nomura,
Nucl. Phys. B \textbf{964} (2021), 115314
doi:10.1016/j.nuclphysb.2021.115314
[arXiv:2003.07638 [hep-ph]].



\bibitem{Matsui:2021khj}
T.~Matsui, T.~Nomura and K.~Yagyu,
Nucl. Phys. B \textbf{971} (2021), 115523
doi:10.1016/j.nuclphysb.2021.115523
[arXiv:2102.09247 [hep-ph]].


\bibitem{Kang:2021jmi}
D.~W.~Kang, J.~Kim and H.~Okada,
Phys. Lett. B \textbf{822} (2021), 136666
doi:10.1016/j.physletb.2021.136666
[arXiv:2107.09960 [hep-ph]].


\bibitem{Bhatia:2021eco}
D.~Bhatia, N.~Desai and A.~Dighe,
JHEP \textbf{04} (2022), 163
doi:10.1007/JHEP04(2022)163
[arXiv:2109.07093 [hep-ph]].






\bibitem{KamLAND-Zen:2022tow}
S.~Abe \textit{et al.} [KamLAND-Zen],
Phys. Rev. Lett. \textbf{130} (2023) no.5, 051801
doi:10.1103/PhysRevLett.130.051801
[arXiv:2203.02139 [hep-ex]].



\bibitem{Esteban:2020cvm}
I.~Esteban, M.~C.~Gonzalez-Garcia, M.~Maltoni, T.~Schwetz and A.~Zhou,
JHEP \textbf{09} (2020), 178
doi:10.1007/JHEP09(2020)178
[arXiv:2007.14792 [hep-ph]]. NuFIT 5.2 (2022), www.nu-fit.org

\bibitem{Planck:2018vyg}
N.~Aghanim \textit{et al.} [Planck],
Astron. Astrophys. \textbf{641} (2020), A6
[erratum: Astron. Astrophys. \textbf{652} (2021), C4]
doi:10.1051/0004-6361/201833910
[arXiv:1807.06209 [astro-ph.CO]].


\bibitem{ATLAS:2019erb}
G.~Aad \textit{et al.} [ATLAS],
Phys. Lett. B \textbf{796} (2019), 68-87
doi:10.1016/j.physletb.2019.07.016
[arXiv:1903.06248 [hep-ex]].


\bibitem{CMS:2021ctt}
A.~M.~Sirunyan \textit{et al.} [CMS],
JHEP \textbf{07} (2021), 208
doi:10.1007/JHEP07(2021)208
[arXiv:2103.02708 [hep-ex]].


   
\end{thebibliography}
\end{document}